\documentclass[aps,prb,twocolumn,reprint,groupedaddress,showpacs]{revtex4-1}
\usepackage{graphicx}
\usepackage{verbatim}
\usepackage{amssymb,amsmath}
\usepackage{float}
\usepackage{CJK}
\usepackage{eqnarray}
\usepackage{sidecap}
\usepackage{enumerate}
\sidecaptionvpos{figure}{c}

\newcommand{\ket}[1]{| #1\rangle}
\newcommand{\bra}[1]{\langle #1 |}
\newcommand{\braket}[2]{\langle #1 | #2\rangle}
\newcommand{\abs}[1]{| #1 |}

\begin{document}


\title{Probability distributions of continuous measurement results for two non-commuting variables and conditioned quantum evolution}


\author{A. Franquet}
\email[]{A.FranquetGonzalez@tudelft.nl}

\author{Yuli V. Nazarov}
\affiliation{Kavli Institute of Nanoscience, Delft University of Technology, 2628 CJ Delft, The Netherlands}


\date{\today}

\begin{abstract}
We address the statistics of a simultaneous CWLM of two non-commuting variables on a few-state quantum system subject to a conditioned evolution. Both conditioned quantum measurement and that of two non-commuting variables differ drastically for either classical or quantum projective measurement, and we explore the peculiarities brought by the combination of the two.

 We put forward a proper formalism for the evaluation of the distributions of measurement outcomes. We compute and discuss the statistics in idealized and experimentally relevant setups. We demonstrate the visibility and manifestations of the interference between initial and final states in the statistics of measurement outcomes for both variables in various regimes. 
  We analytically predict the peculiarities at the circle ${\cal O}^2_1+{\cal O}^2_2=1$ in the distribution of measurement outcomes in the limit of short measurement times and confirm this by numerical calculation at longer measurement times.  We demonstrate analytically anomalously large values of the time-integrated output cumulants in the limit of short measurement times(sudden jump) and zero overlap between initial and final states, and give the detailed distributions. We present the numerical evaluation of the probability distributions for experimentally relevant parameters in several regimes and demonstrate that interference effects in the conditioned measurement can be accurately predicted even if they are small.
\end{abstract}

\maketitle

\section{Introduction}
\label{sec:intro}
\label{sec:intro}
Quantum computing and communication\cite{NielsenChuang} are stimulating rapid progress in the understanding and control of small quantum systems. An important ingredient for advanced quantum control is the ability to realize continuous monitoring of a quantum system. Theories for continuous quantum measurement\cite{CWLM0,CWLM1,CWLM2,CWLM25,NazWei,CWLM3,CWLM4} and experiments\cite{Devoret, SiddiqiSingle, SiddiqiEntanglement,Huard, DiCarlo, SiddiqiMapping, SiddiqiMolmer} have enabled a detailed understanding of the realisitic and practical measurement process in quantum mechanics.\\
A relevant case of quantum measurement is that of the measurement of non-commuting quantum variables. The fact that some observables {\it cannot} be measured together is one of the major differences between quantum and classical theory. Although it is possible to perform a simultaneous measurement of non-commuting variables, only recently\cite{CWLM2, NazWei, MolmerMonitoring, ProbingQ, SiddiqiDynamics, SiddiqiKorotkovCorrelators} the dynamics of the qubit state has been studied under these conditions. It is important to note that the simultaneous measurement of non-commuting variables for a long time has been a topic of many experimental and theoretical studies in quantum optics\cite{SimOptics}. The linearity of most optical measurements provides a perfect platform for experiments of this kind.

Another interesting and relevant kind of quantum measurement is the conditioned measurement. For a general conditioned evolution, both the initial and final states of the system can be regarded as fixed. This is achieved by the selection of the measurement results on the basis of the result of the concluding projective measurement.  It has been shown that the statistics of such a conditioned measurement may drastically differ from the unconditioned case\cite{CWLM25, NazWei}. In this context, the statistics of measurement results reveal purely quantum phenomena that can be interpreted in terms of weak values\cite{BigWeakValues} and associated with the interference of initial and final states\cite{WisemanWeakValues, PastQStateMolmer} .\\

This paper elaborates on a combined case of quantum measurement of two non-commuting variables and for conditioned quantum evolution. The goal is to inspect the full statistics of the measurement results and its dependence on the dynamics of the system measured. For that purpose,  we use the theory of continuous weak linear measurements (CWLM), where a sufficiently weak coupling between a quantum system and infinitely many degrees of freedom of a linear detector provides their entanglement and conversion of the (discrete) quantum information into continuous time-dependent readings of the detector\cite{CWLM0,CWLM1,CWLM2,CWLM25,NazWei,CWLM3,CWLM4}. Our approach to CWLM statistics was first introduced in \cite{NazKin, NazWei}, and extended to include conditioned evolution in \cite{FranquetNaz}. It is based on the theory of full counting statistics in the extended Keldysh formalism\cite{AQMbook}.The statistics of measurements of $\int dt \hat{V}(t)$, $V(t)$ being a quantum mechanical variable representing linear degrees of freedom that are measured,  are evaluated with the characteristic functional method and the use of counting field technique. The method provides the necessary and compact description of the whole quantum system consisting of the measured system and multiple degrees of freedom describing general linear detectors.\\

The probability distributions for the measurement of a single variable have been extensively studied in our recent publication \cite{FranquetNaz}. The motivation to address the two-variable case comes from the recent experiments\cite{Huard} where a qubit has been measured in a resonance fluorescence setup. In the experiment, the transmon qubit enclosed in a non-resonant three-dimensional (3D) superconducting cavity is resonantly driven at the Rabi frequency $\Omega$ and its fluorescence signal is recorded.  The cavity is coupled to two transmission lines, the resonant driving field drives the qubit via a weakly coupled line, while most of the fluorescence signal exits via the other line which is coupled strongly. The complex amplitude of the the fluorescence signal is proportional to $\sigma_{-}$, the average of the lowering operator $\hat{\sigma}_{-}=\ket{g}\bra{e}$ of the qubit, and oscillates with the Rabi frequency $\Omega$. At each run, the qubit is prepared in either its ground $\ket{g}$ or excited $\ket{e}$ state and the signal is monitored during a time interval $(0,{\cal T})$. At time ${\cal T}$, the qubit is projectively measured using a strong pulse at the bare cavity frequency.\\
A heterodyne detection setup is used to measure this signal, and the fluorescence signal can be interpreted as a result of a weak continuous measurement. We notice that the experiment discussed can give access not only to the conditioned averages, but also to the conditioned statistics of the measurement results. Those are statistics of the continuous weak measurement of two non-commuting variables of the qubit, $\sigma_{x}$ and $\sigma_{y}$ that comprise $\hat{\sigma}_{-} = \hat{\sigma}_{x} - i\hat{\sigma}_{y}$.

The statistics of the conditioned measurement results reveal the signatures of interference between pre and post selected states. With the present work, we extend these signatures to the case of simultaneous measurement of non-commuting variables, and reveal the relation between the visibility of these signatures and the qubit dynamics in different parameter regimes.

Our results demonstrate that one can achieve very detailed theoretical predictions of the statistics of CWLM of two non-commuting variables, with account for every relevant experimental parameter. This allows for the study and characterization of quantum effects at any choice of parameters, even in the regime where the signatures are very weak.

Among other interesting results, we show that the joint distribution of measurement outcomes of two non-commuting quantum variables $P({\cal O}_1,{\cal O}_2)$ has peculiarities located at the circle ${\cal O}_1^2 + {\cal O}_2^2 =1$. This is the two-variable analog of the {\it half-quantized} measurement values for the single variable measurement case. 
We reveal these peculiarities by analytical calculation of the quasi-distribution of shifts in the limit of short measurement time, and demonstrate them in numerical results at larger measurement times.
We demonstrate how the visibility of the circle is suppressed by the system dynamics, such that the joint distribution effectively becomes a product of two independent distributions $P({\cal O}_1,{\cal O}_2)\approx P_1({\cal O}_1)P_2({\cal O}_2)$.\\

At measurement times that are so short that the wave function of the system does not change significantly, and in the case of zero or small overlap between initial and final states, we reveal anomalously large values of the cumulants of the distribution function of time integrated outputs that we previously nicknamed {\it sudden jump} \cite{FranquetNaz}. In the case of simultaneous measurement of two non-commuting variables, we reveal simultaneous {\it sudden jump} of the two time integrated outputs ${\cal O}_1, {\cal O}_2$ with an appropriate choice of Hamiltonian. For the average value of the output, the big values are readily understood from the weak value theory\cite{BigWeakValues}. We present both analytical and numerical results.

We also compute the distributions of the outputs under realistic experimental parameters of \cite{Huard} concentrating on the quantum signatures of conditioned evolution and the non-commutativity of the variables.

The structure of the paper is as follows. We outline and develop the necessary formalism in Sec. \ref{sec:method}, starting from a Bloch master equation for the qubit evolution that is augmented  with counting fields to describe the statistics of detector readouts. This formalism has been elaborated in our previous work\cite{FranquetNaz}, we extend it here  to the simultaneous measurement of two non-commuting variables. We reveal the role of various experimental parameters and formulate the relevant quantum noise inequalities for a general multiple detector setup. In Sec. \ref{sec:quasidistributions} we concentrate on short ${\cal T}$  and compute the quasi-distribution of the shifts of the joint distribution $P({\cal O}_1,{\cal O}_2)$, revealing the circle shape discussed. In Sec. \ref{sec:smalltimes} we concentrate on the case of zero overlap between initial and final states and derive analytical expressions for the joint distribution $P({\cal O}_1,{\cal O}_2)$ of measurement outcomes at short times. In this regime, the joint distribution is essentially non-Gaussian and manifests the {\it sudden jumps} in the integrated outputs. 

In the next sections, we present numerical results at various times scales and in parameter regimes demonstrating the possibility of very detailed predictions of CWLM distributions. To start with, in Sec. \ref{sec:NumShort} we present numerical simulations at time intervals that are much smaller than the typical time scales of all Hamiltonian dynamics focusing on three relevant cases: the case of ideal detectors, and the experimentally relevant case with and without detuning. In Sec. \ref{sec:NumLong}, we concentrate on time scales of the order of the decoherence time, inspecting the three cases for ideal detectors with and without drive, and for experimentally relevant setup. We conclude in Sec. \ref{sec:conclusion}.\\

\section{Method}
\label{sec:method}
The description of CWLM in use was first introduced in\cite{NazKin}, and later, extended in \cite{FranquetNaz} to compute probability distributions of a continuous measurement for a conditioned quantum evolution.\\
In contrast to other methods such as path integral formulation\cite{NazWei,CWLM3}, effective action method\cite{CWLM0,CWLM4}, past states formalism\cite{PastQStateMolmer} or the stochastic update equation\cite{trajectory}; this description permits the direct evaluation of the generating function of the probability distribution of the measurement results.\\

The central object in this description is a Bloch-master equation for the evolution of the quasi-density matrix of the quantum system that is augmented with counting fields. Evaluating the trace of the augmented density matrix from this equation as a function of the counting fields provides the generating function for the probability distribution of the detector(s) output(s). We give the concrete expression of such equation for a simultaneous measurement of two variables ${\cal O}_{1}, {\cal O}_{2}$ of the quantum system. In an ideal measurement, where all decoherence is due to the coupling with  ${\cal O}_{1}, {\cal O}_{2}$ and for the case of independent detectors, it reads,

\begin{eqnarray}
\label{eq1}
\frac{\partial\hat{\rho}}{\partial t} =& -\frac{i}{\hbar}[\hat{H}_{q},\hat{\rho}] - \sum_{i}\frac{S^{(i,i)}_{QQ}}{\hbar^{2}}\mathcal{D}[\hat{\cal O}_{i}]\hat{\rho} -\frac{\chi_{i}^{2}(t)}{2}S^{(i,i)}_{VV}\hat{\rho} \\*
\nonumber
& -\frac{S^{(i,i)}_{QV}}{\hbar}\chi_{i}(t)[\hat{\rho},\hat{\cal O}_{i}] + \frac{i a^{i,i}_{VQ}\chi_{i}(t)}{2}[\hat{\rho},\hat{\cal O}_{i}]_{+}.
\end{eqnarray}

Here, $[,]$ and $[,]_{+}$ refer to commutator and anti-commutator, respectively, $\mathcal{D}[\hat{A}]\hat{\rho}\equiv\left(\frac{1}{2}[\hat{A}^{\dagger}\hat{A},\hat{\rho}]_{+}-\hat{A}\hat{\rho}\hat{A}^{\dagger}\right)$ and $i=1,2$.\\
For each output ${\cal O}_{i}$, there is a corresponding counting field $\chi_{i}(t)$ and a pair of input $\hat{Q}_{i}$ - output $\hat{V}_{i}$ operators of the corresponding detector. The parameters in the previous equation are the two-point correlators of these input-output operators, that give the set of noises and response functions in this linear measurement environment,

\begin{subequations}
\label{corr}
\begin{eqnarray}
S^{(i,j)}_{QQ} =& \frac{1}{2}{\displaystyle \int_{-\infty}^{t}}dt'\left\langle\left\langle\hat{Q}_{i}(t)\hat{Q}_{j}(t')+\hat{Q}_{j}(t')\hat{Q}_{i}(t)\right\rangle\right\rangle,\\*
S^{(i,j)}_{QV} =& \frac{1}{2}{\displaystyle \int_{-\infty}^{t}}dt'\left\langle\left\langle\hat{Q}_{i}(t)\hat{V}_{j}(t')+\hat{V}_{j}(t')\hat{Q}_{i}(t)\right\rangle\right\rangle,\\*
S^{(i,j)}_{VV} =& \frac{1}{2}{\displaystyle \int_{-\infty}^{t}}dt'\left\langle\left\langle\hat{V}_{i}(t)\hat{V}_{j}(t')+\hat{V}_{j}(t')\hat{V}_{i}(t)\right\rangle\right\rangle,\\*
a^{(i,j)}_{VQ} =& -\frac{i}{\hbar}{\displaystyle \int_{-\infty}^{t}}dt'\left\langle[\hat{V}_{i}(t),\hat{Q}_{j}(t')]\right\rangle,\\*
a^{(i,j)}_{QV} =& -\frac{i}{\hbar}{\displaystyle \int_{-\infty}^{t}}dt'\left\langle[\hat{Q}_{i}(t),\hat{V}_{j}(t')]\right\rangle.
\end{eqnarray}
\end{subequations}

This set of noise and response functions define completely the characteristics of the measurement process. Conforming to the assumption of slow qubit dynamics, the noises are white and the responses are instant, corresponding to zero-frequency correlators.\\
The values of these noises and responses are restricted by a set of Cauchy-Schwartz inequalities of the form, \citep{QNoise}

\begin{equation}
\label{ineq}
S^{(i,i)}_{QQ}S^{(j,j)}_{VV}-\abs{S^{(i,j)}_{QV}}^{2}\geq\frac{\hbar^{2}}{4}\abs{a^{(j,i)}_{VQ}-a^{(i,j)}_{QV}}^{2},
\end{equation}

for each pair of operators $\hat{Q}_{i}$, $\hat{V}_{i}$; and not excluding inequalities for pairs of only input ($\hat{Q}$'s) or only output ($\hat{V}$'s) operators.

As discussed in\cite{FranquetNaz}, these inequalities impose the necessary conditions for the positivity of the probability distributions of measurement outputs. However, it is possible an necessary to derive a more restrictive set of inequalities that impose the conditions for this positivity. In two-detector case at hand, an extra restriction reads:

\begin{align}
\label{ineq2}
\nonumber
S^{(1,1)}_{QQ} + S^{(2,2)}_{QQ} \geq \frac{\hbar^{2}}{4}\frac{\abs{a^{(1,1)}_{VQ}-a^{(1,1)}_{QV}}^{2}}{S^{(1,1)}_{VV}}+\frac{\abs{S^{(1,1)}_{QV}}^{2}}{S^{(1,1)}_{VV}}\\ 
\nonumber
+ \frac{\hbar^{2}}{4}\frac{\abs{a^{(2,2)}_{VQ}-a^{(2,2)}_{QV}}^{2}}{S^{(2,2)}_{VV}}+\frac{\abs{S^{(2,2)}_{QV}}^{2}}{S^{(2,2)}_{VV}}\\
\nonumber
+\hbar\biggl|\frac{\left(a^{(1,1)}_{VQ}-a^{(1,1)}_{QV}\right)S^{(2,1)}_{QV}}{S^{(1,1)}_{VV}}-\frac{\left(a^{(2,2)}_{VQ}-a^{(2,2)}_{QV}\right)S^{(1,2)}_{QV}}{S^{(2,2)}_{VV}}\biggr|\\
+\frac{\abs{S^{(2,1)}_{QV}}^{2}}{S^{(1,1)}_{VV}} + \frac{\abs{S^{(1,2)}_{QV}}^{2}}{S^{(2,2)}_{VV}} .
\end{align}

We demonstrate in Appendix \ref{sec:Positivity} how to derive such inequalities from analytical expressions of the joint distribution of measurement outcomes.
Those and more complex inequalities can be derived from the positivity of the matrix
$S_{\beta\alpha}  +i \frac{a_{\beta \alpha}-a_{\alpha \beta}}{2}$ where the indices $\alpha,\beta$ index the whole set of operators $\hat{V},\hat{Q}$.
\\

Let us focus in a experimental situation general to the one described in\cite{Huard}, a transmon qubit embedded in a 3D superconducting cavity with two levels split in $z$-direction under conditions of strong resonant drive that compensates the splitting of the qubit levels. The effective  Hamiltonian reads
\begin{equation}
\hat{H}_{q}=\frac{\hbar}{2}\Omega\hat{\sigma}_{x}+ \frac{\hbar}{2}\Delta\hat{\sigma}_{z},
\end{equation}
$\Omega$ being the Rabi frequency proportional to the amplitude of the resonant drive, and $\Delta$ being the detuning of the drive frequency from the qubit energy splitting. The interaction with the environment induces decoherence, excitation and relaxation of the qubit, with the rates $\gamma_{d},\gamma_{\uparrow},\gamma_{\downarrow}$ respectively. The measured quantities are related to the fluorescence signal emitted from the qubit, so ${\cal O}_1$ and ${\cal O}_2$ are conveniently chosen to be $\sigma_x$ and $\sigma_y$. \\

This is the case of heterodyne detection. The signal from $\sigma_{x,y}$ eventually oscillates at frequency $\Omega$. The accumulating signal is obtained by the mixture of this signal with the resonant drive. As a result, it is in principle possible to measure both $\sigma_{x,y}$ signals with a single detector variable mixing it with $\sin$ and $\cos$ components of the resonant drive. Then Eq. \ref{eq1} needs to be adjusted to the case of heterodyne detection. The symmetrized noises $S_{VV}$ have to be taken at frequency $\Omega$ rather than on zero frequency. The same pertains the susceptibilities. The most important change concerns the second term in Eq. \ref{eq1} that, for ${\cal O}_{1,2} = \sigma_{x,y}$ describes the decoherence and transitions between the states $\sigma_z\ket{Z^{\pm}} = \pm \ket{Z^{\pm}}$. In Eq. \ref{eq1}, the rates of these transitions are equal for both directions, $\gamma_{\downarrow} = \gamma_{\uparrow}$. For the case of heterodyne detection, they are not: there are two rates with gaining/loosing energy proportional to the quantum noise $S_{QQ}$ at positive and negative frequencies $\pm \Omega$. We also need to add the terms describing the decoherence of the states $\ket{Z^{\pm}}$.

With this, the equation reads, ($i=1,2$)
\begin{eqnarray}
\label{eq1exp}
\frac{\partial\hat{\rho}}{\partial t} =& -\frac{i}{\hbar}[\hat{H}_{q},\hat{\rho}] -\gamma_{d}\mathcal{D}[\hat{\sigma}_{z}]\hat{\rho}-\gamma_{\uparrow}\mathcal{D}[\hat{\sigma}_{+}]\hat{\rho}\\*
\nonumber
&-\gamma_{\downarrow}\mathcal{D}[\hat{\sigma}_{-}]\hat{\rho} -\frac{S_{QV}}{\hbar}\sum_{i}\chi_{i}(t)[\hat{\rho},\hat{{\cal O}_{i}}]\\*
\nonumber
& + \frac{i a_{VQ}}{2}\sum_{i}\chi_{i}(t)[\hat{\rho},\hat{{\cal O}_{i}}]_{+}-\sum_{i}\frac{\chi_{i}^{2}(t)}{2}S_{VV}\hat{\rho},
\end{eqnarray}
$\hat{\sigma}_{+}$ ($\hat{\sigma}_{-}$) being the rising and lowering operators of the qubit, and $\hat{\sigma}_{z}=\ket{e}\bra{e}-\ket{g}\bra{g}$ the standard Pauli operator.\\
All the parameters entering the equation can be characterized from experimental measurements. We provide an example of concrete values in Section \ref{sec:NumShort}.\\

For simplicity we inspect the case of identical but independent detectors. Meaning all cross noises and responses vanish and the behavior of both detectors is physically the same. In that case, the rates and noises are restricted by the inequality,  

\begin{equation}
\label{ineqideal}
S_{QQ}S_{VV}-\abs{S_{QV}}^{2}\geq\frac{\hbar^{2}}{4}\abs{a_{VQ}-a_{QV}}^{2}.
\end{equation}

For a simple system like a single qubit it is natural to make the measured operator dimensionless, with eigenvalues of the order of one, or, even better, $\pm 1$. With this, one can define and relate the measurement induced dephasing rate $2 \gamma = 2S_{QQ}/\hbar^{2}$ and the acquisition time $t_{a}\equiv 4 S_{VV}/\abs{a_{VQ}}^{2}$ required to measure the variable ${\cal O}_{1,2}$ with a relative accuracy $\simeq 1$.\\

We concentrate on the simultaneous measurement of two variables of a qubit during a time interval $(0,{\cal T})$. During this time interval, one accumulates the time-dependent outputs of the detectors and normalize them by the same interval, $V_{i} \equiv \frac{1}{{\cal T}}\int_{0}^{{\cal T}} V_{i}(t') dt'$ ($i=1,2$). Our goal is to evaluate the joint probability distribution $P(V_{1}, V_{2})$ of the measurement results, conditioned to an initial qubit state given by $\hat{\rho}(0)$, and to a post-selection of the qubit in a specific state $\ket{\Psi}$ at the time moment ${\cal T}$. This involves the projection on the state $\ket{\Psi}$, represented by the projection operator $\hat{P}_{\Psi}=\ket{\Psi}\bra{\Psi}$ . This works under assumption of an ideal and fast post-selection so that the system measured is projected on a known pure state $|\Psi\rangle$. This is the case of the experimental setup \cite{Huard}. In reality, there can be errors in the post-selection.  Such errors can also be accounted for in the formalism outlined. To this end, one replaces the projection operator $\hat{P}_{\Psi}$ with a density matrix-like Hermitian operator $\hat{\rho}_f$ satisfying ${\rm Tr } [\hat{\rho}_f]  = 1$. For instance, if after a faulty projection measurement with the result "1" the system is in a orthogonal state $|\Psi_2\rangle$ with probability $p_e$, the corresponding $\hat{\rho}_f$ reads
\begin{equation}
\hat{\rho}_f = (1-p_e) |\Psi_1\rangle\langle\Psi_1| + p_e |\Psi_2\rangle\langle\Psi_2| 
\end{equation}

The probability distribution of the detector outcomes can be computed from the generating function according to

\begin{equation}
\label{eq2}
P(V_{1},V_{2}) = \frac{{\cal T}}{2\pi}\int\int d\chi_{1}d\chi_{2} e^{-i\chi_{1} V_{1}{\cal T}}e^{-i\chi_{2} V_{2}{\cal T}}C(\chi_{1},\chi_{2}; {\cal T}).
\end{equation}

The joint statistics are extracted from the quasi-density matrix $\hat{\rho}(\chi_{1},\chi_{2}; {\cal T})$ at the end of the interval calculated using Eq. \eqref{eq1}. With the post-selection, the quasi-density matrix is projected on the final state measured $\ket{\Psi}$, and the conditioned generating function of the detector outcomes reads \cite{FranquetNaz}

\begin{equation}
\label{eq3}
\tilde{C}(\chi_{1},\chi_{2}; {\cal T}) = \frac{\text{Tr}_{q}(\hat{P}_{\Psi}\hat{\rho}(\chi_{1},\chi_{2}; {\cal T}))}{\text{Tr}_{q}(\hat{P}_{\Psi}\hat{\rho}(\chi_{1}=0,\chi_{2}=0; {\cal T}))}.
\end{equation}
Here, $\text{Tr}_{q}$ denotes the trace over qubit variables.\\
Sometimes it is convenient to normalize the time-integrated outputs introducing ${\cal O}_{i} = V_{i}/a^{(i,i)}_{VQ}$ that immediately corresponds to the eigenvalues of $\hat{\cal O}_{i}$ (We stress that ${\cal O}_{i}$ are coming from the averaging of an environmental operator rather than $\hat{\cal O}_{i}$).

\section{Quasi-distribution of shifts}
\label{sec:quasidistributions}

For a sufficiently long measurement, the distribution of the measurement results is a shifted Gaussian with the value of the shift proportional to the averaged value of the operator measured. In this case, the spread of the Gaussian is much smaller than the shift. In this Section, we will attempt to understand the shifts in the limit of short measurement times ${\cal T}$. In principle, 
any distribution of a vector variable $P(\vec{{\cal O}})$ can be presented as a convolution of a Gaussian distribution $P_G$ and a quasi-distribution of the shifts,
\begin{equation}
P(\vec{{\cal O}}) = \int d\vec{s} C(\vec{s}) P_G(\vec{{\cal O}} -\vec{s})
\end{equation}
One should only not to be confused with the fact that $С$ is a quasi-distribution and should not be ever positively defined.

The convolution of such kind is especially natural since the solution of Eq. \eqref{eq1} is proportional to the characteristic function of the Gaussian distribution. If we neglect the cross-noises, and the Hamiltonian dynamics, the solution at short ${\cal T}$ can be represented as
\begin{equation}
\hat{\rho}({\cal T}) = \exp\left(-\frac{{\cal T}\chi_{i}^{2}(t)}{2}S^{(i,i)}_{VV}\right) \hat{U} \hat{\rho}(0) \hat{U}
\end{equation}
with $\hat{U} = \exp\left(\frac{i a^{i,i}_{VQ}\chi_{i}{\cal T}}{2}\right)$
The first factor here is the characteristic function of the Gaussian distribution generated by the detector noises. From the second factor, assuming the initial density matrix $\hat{\rho}_i$ and the post-selection described by $\hat{\rho}_f$, we obtain 
the generating function of the shift quasi-distibution 
\begin{equation}
C(\vec{\chi})=\frac{{\rm Tr}[\hat{\rho}_f \hat{U}\hat{\rho}_i\hat{U}]}{{\rm Tr}[\hat{\rho}_f \hat{\rho}_i]}.
\end{equation}

We illustrate the quasi-distribution of the shifts for the case of a qubit. Although in this paper we concentrate on two-detector setups, it is much more instructive to consider now {\it three } detectors measuring all three Pauli matrices $\vec{\sigma} = (\sigma_x,\sigma_y, \sigma_z)$.
We normalize the detector outputs on $\pm1$ of  Pauli matrix eigenvalues and rescale the corresponding counting fields $\vec{\chi}$. accordingly. With this, the matrix $\hat{U}$ becomes nicely symmetric
\begin{equation}
\hat{U} = \exp\left(-i {(\vec{\chi},\vec{\sigma})}/{2}\right)
\end{equation}

The final and initial density matrices for a qubit are represented as 
\begin{equation}
\hat{\rho}_{i,f} = \frac{1}{2}(1+ (\vec{P}_{i,f},\vec{\sigma}))
\end{equation}
with polarization vectors $|\vec{P}_{i,f}|<1$.
The generation function for smaller number of detectors is obtained by setting some components of $\vec{\chi}$ to $0$. For instance, setting $\chi_{z,y}=0$ gives
\begin{eqnarray}
C(\chi_x) =& \left(1+(\vec{P}_i,\vec{P}_f)\right)^{-1}\biggl(((\vec{P}_i,\vec{P}_f) -P_i^z P_f^z)\nonumber \\
&+(1+P_i^z P_f^z)\cos \chi_x +(P_i^z +P_f^z)\sin\chi \biggr)
\end{eqnarray}
which corresponds to the following quasi-distibution of the shifts
\begin{equation}
C(s_x) = \delta(s_x-1) + \delta(s_x) + \delta(s_x+1)
\label{eq:shifts1d}
\end{equation}
This quasi-distribution, as discussed in \cite{FranquetNaz}, is located on a compact support of half-sums of the eigenvalues $\pm 1$ of the operator $\sigma_x$. The half-quantized value $s_x =0$ is manifested only in the case of conditioned measurements.

Multiplying the matrices and taking the trace, we obtain the answer for three detectors. It can be naturally separated into scalar, vector, and tensor part ($\chi \equiv |\vec{\chi}|$),
\begin{eqnarray}
C_s(\vec{\chi})&=& \cos\chi +  (\vec{P}_i,\vec{P}_f) \\
C_v(\vec{\chi})&=& i (\vec{P}_i+\vec{P}_f,\chi) \frac{\sin\xi}{\xi}\\
C_t(\vec{\chi})&=& - (\vec{P}_i,\vec{\chi})(\vec{P}_f,\vec{\chi})\frac{2 \sin^2 (\chi/2)}{\chi^2}\\
C&=&\frac{C_s + C_v +C_t}
{1+(\vec{P}_i,\vec{P}_f)}.
\end{eqnarray}

Let us now compute the quasi-distribution of the shifts the inverse Fourier transform of $C$,
\begin{equation}
C(\vec{s}) = \int \frac{d \vec{s}}{(2\pi)^3} C(\vec{\chi}) \exp(- i (\vec{s},\vec{\chi}))
\end{equation}
Eventually, the integral is rather involved. The best way to perform the integration is to try the direct transform. We  note that
\begin{equation}
\frac{\sin(\chi A)}{\chi} \equiv z(A)=\int d \vec{s} \frac{\delta(s-A)}{4\pi A} \exp(i (\vec{s},\vec{\chi}))
\end{equation}
at any $A$ and  
\begin{eqnarray}
\frac{\sin(\chi)}{\chi}, \cos(\xi) =\lim_{\mathop{A \to 1}}  z(A), \frac{d}{d A} z\;;\\
\frac{2 \sin^2(\chi/2)}{\chi^2} = \int_0^1 {d A} z(A)
\end{eqnarray}
With using this we arrive at the quasi-distribution of the form 
\begin{eqnarray}
\mkern-18mu C_s(\vec{s})&=& -\frac{1}{4\pi} (\delta(s-1)+\delta'(s-1))+  (\vec{P}_i,\vec{P}_f) \delta(\vec{n})\\
\mkern-18mu C_v(\vec{n})&=& -(\vec{P}_i+\vec{P}_f,\frac{\partial}{\partial\vec{s}}) \delta(s-1)\\
\mkern-18mu C_t(\vec{n})&=&  (\vec{P}_i,\frac{\partial}{\partial\vec{s}})(\vec{P}_f,\frac{\partial}{\partial\vec{s}})\frac{\Theta(1-s)}{s} \\
\mkern-18mu C&=&\frac{C_s + C_v +C_t}{1+(\vec{P}_i,\vec{P}_f)}
\label{eq:shifts3d}
\end{eqnarray}

We observe that the vector and tensor contributions provide a quasi-distribution located on a compact support
$s=0$ or $s=1$. The latter is rather surprising: it invokes a notion of a 'classical' qubit spin, a classical unit vector pointing in an arbitrary direction. While for such classical spin the quasi-distribution would have been positive, this is not the case of actual quantum mechanical expression: the quasi-distribution is made of $\delta$-function and its derivatives. We do not find it instructive to plot the resulting quasi-distribution.
The tensor part also contains terms located on this support. In addition, there are terms $\propto (\vec{P}_f,\vec{s})(\vec{P}_i,\vec{s})/s^5$ located within the sphere $s<1$. The tensor part persists only for the case of conditional measurement $\vec{P}_f \ne 0$.

To obtain the distribution of 2 outputs, we integrate it over $s_z$ making use of 
\begin{equation} 
\int d s_z \ z(A)= \frac{2}{\sqrt{A^2 - s_\perp^2}}; \; s_\perp \equiv \sqrt{s_x^2+s_y^2}
\end{equation}

The resulting quasi-distribution reads (here, the indices $a,b = x,y$)

\begin{eqnarray}
C_s(\vec{s})&=& -\frac{1}{2\pi} \frac{1}{(1-s^2_\perp)^{3/2}}+  (\vec{P}_i,\vec{P}_f) \delta(s_x)\delta(s_y)\\
C_v(\vec{s})&=& -({P}^a_i+{P}^a_f,\frac{\partial}{\partial n^a}) \delta(s-1)\\
C_t(\vec{s})&=&  {P}^a_i\frac{\partial}{\partial{s^a}}P^b_f\frac{\partial}{\partial s^b}){\rm arccosh} (s^{-1}_\perp)\\
C&=&\frac{C_s + C_v +C_t}{1+(\vec{P}_i,\vec{P}_f)} \label{eq:shifts2d}
\end{eqnarray}

We see that this quasi-distribution is located at the compact support $s_x^2+s_y^2=1, s_x^2+s_y^2=0$ as well as inside the circle $s_x^2+s_y^2<1$. This gives us an expectation that the actual distribution of the measurement results should exhibit some peculiarities at $s_x^2+s_y^2=1$, an expectation that is confirmed by numerical results of subsequent Sections.

It is worth noting that the generalized functions involved in the quasi-distributions presented in the Eqs. \eqref{eq:shifts2d} and \eqref{eq:shifts3d} are rather involved and should be dealt with carefully. In particular, a direct attempt to integrate Eq.\eqref{eq:shifts2d} over $n_y$ does not immediately reproduce Eq. \eqref{eq:shifts1d} as it should. Rather, the integration diverges near $s_x^2+s_y^2=1$. To resolve this apparent paradox, one requires a regularization of the generalized functions involved. Such regularization can be provided by replacing
\begin{equation}
\delta(s-A) \to \pi^{-1} {\rm Im} \frac{1}{A + i \xi }
\end{equation}
at small but finite $\xi$. With this, the divergence at the circle edge is eliminated and Eq. \eqref{eq:shifts1d} is reproduced.

\section{Short time intervals and zero overlap}
\label{sec:smalltimes}
In this Section, we consider again very short ${\cal T}$ such that the change of the density matrix due to Hamiltonian and dissipative dynamics is small. Since the measuring time is too short to resolve the signal with sufficient accuracy, we expect the distribution to be close to the Gaussian one 
\begin{equation}
P_G({\cal O}_{1},{\cal O}_{2})= \prod_{\mathop{i=1,2}}\frac{1}{\sigma_i\sqrt{2\pi}}\exp\left(-\frac{{\cal O}_{i}^2}{2\sigma^2_i}\right),
\end{equation}
with $ \sigma^2_i = S^{(i,i)}_{VV}/({\cal T}\abs{a^{(i,i)}_{VQ}}^2)$. The spread of ${\cal O}$ is much larger than their eigenvalues. However, the distribution can become quite different if the overlap between the initial state, $\ket{i}$, and the final state of the projective measurement, $\ket{\Psi}$, vanishes: $\braket{i}{\Psi}=0$. The latter implies that such output of the projective measurement is very improbable, but it can be singled out and its statistics are worth studying.\\

\begin{figure*}[!ht]
\centering
\includegraphics[width=1.1\textwidth, height=5cm]{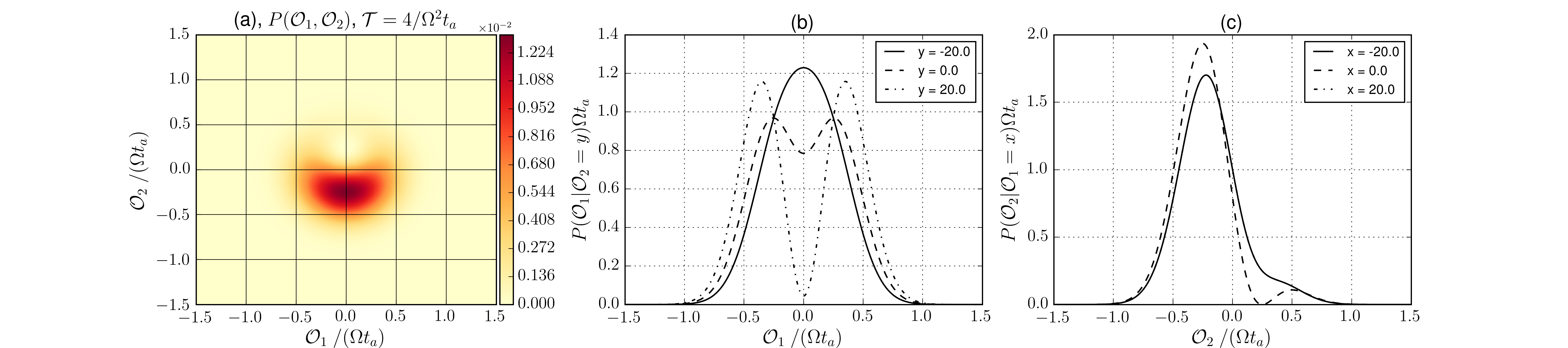}
\caption{(Color online) (a): Probability distribution of outputs [Eq. \eqref{smalltimes}] in the sudden jump regime in case of an ideal detector ($K\equiv\gamma t_a=1$). The figures (b) and (c) present  conditioned distributions. In (b), we plot the probability distribution of ${\cal O}_1$ output given a ${\cal O}_2=y$ result for the other output. (c) gives the probability distribution of ${\cal O}_2$ output given a ${\cal O}_1=x$ result for the other output. All distributions are evaluated at ${\cal T}=4/(\Omega^2t_a)$.}
\label{fig:smallanalytical}
\end{figure*}

\begin{figure*}[!ht]
\centering
\includegraphics[width=1.1\textwidth, height=5cm]{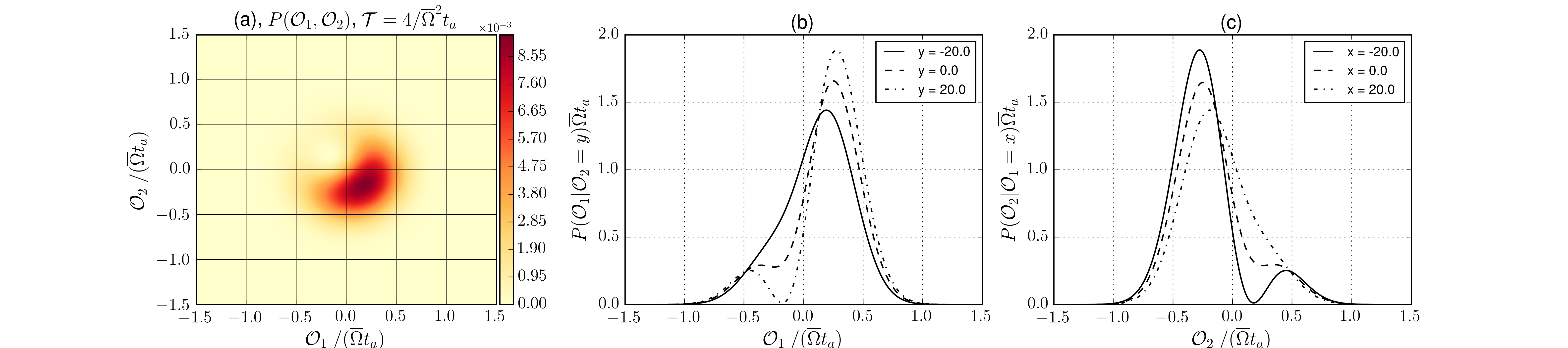}
\caption{(Color online) (a): Probability distribution of outputs [Eq. \eqref{smalltimesxy}] in the sudden jump regime in case of an ideal detector ($K\equiv\gamma t_a=1$). Both figures (b) and (c) present conditioned distributions. (b) gives the probability distribution of ${\cal O}_1$ output given a ${\cal O}_2=y$ result for the other output. (c) gives the probability distribution of ${\cal O}_2$ output given a ${\cal O}_1=x$ result for the other output. All the distributions are evaluated at ${\cal T}=4/(\bar{\Omega}^2t_a)$. We set $\Omega_x=\Omega_y$, this explains the symmetry.}
\label{fig:smallxyanalytical}
\end{figure*}

A peculiarity termed a {\it sudden jump} of the integrated output, is characteristic for this situation. It can be explained from the concept of weak values\cite{WeakValues} as far as average outputs are concerned. For the whole statistics of the outputs, the sudden jump was studied for a single variable measurement\cite{FranquetNaz}. Here we demonstrate that the sudden jump is still seen in the statistics of simultaneous measurement of two variables. A proper choice of Hamiltonian permits for a simultaneous sudden jump in both integrated outputs. The signature of sudden jump is enhanced in the distribution where the distribution of one output is conditioned on a specific value of another one.\\

To this end, let us focus first at the experimental situation in \cite{Huard}. In this case, the two measured variables are conveniently chosen to be $\hat{\sigma}_x$ and $\hat{\sigma}_y$ and the qubit is driven by a Hamiltonian $\hat{H}_q= \hbar\frac{\Omega}{2}\hat{\sigma}_x$. In the simplest case where the two detectors are independent and no cross noises are present, and with the assumptions of short ${\cal T}$ and zero overlap $\braket{i}{\Psi}=0$ (the qubit is prepared in $\ket{Z^{+}}$ and post-selected in $\ket{Z^{-}}$), one obtains the following joint characteristic function of the distribution of detector outputs:\\

\begin{widetext}

\begin{equation}
C(\chi, {\cal T}) = \frac{4\gamma + {\cal T}\left(\left(\Omega - ia^{(2,2)}_{VQ}\chi_2\right)^2 - \left(a^{(1,1)}_{VQ}\chi_1\right)^2\right)}{4\gamma + {\cal T}\Omega^2}e^{-\frac{1}{2}\sum_iS^{(i,i)}_{VV}\chi^2_i{\cal T}}.
\end{equation}

\end{widetext}
 $\gamma= S^{(1,1)}_{QQ}/\hbar^2+S^{(2,2)}_{QQ}/\hbar^2$ being the  decoherence rate.\\
This gives the average outputs

\begin{equation}
\begin{split}
\bar{\cal O}_1 = 0,
\end{split}
\quad\quad\quad\quad
\begin{split}
\bar{\cal O}_2 = -\frac{2\Omega}{4\gamma + {\cal T}\Omega^2};
\end{split}
\end{equation}

and the joint distribution
\begin{widetext}

\begin{equation}
\label{smalltimes}
P({\cal O}_1, {\cal O}_2) = \frac{1}{4\gamma + {\cal T}\Omega^2}\left(4\gamma + {\cal T}\left[\left(\Omega - \frac{4{\cal O}_2}{t_{a2}}\right)^2 - \frac{4}{{\cal T}t_{a2}}+\frac{4}{t_{a1}}\left(\frac{4{\cal O}^2_1}{t_{a1}} - \frac{1}{\cal T}\right)\right]\right) P_G({\cal O}_1, {\cal O}_2)
\end{equation}

\end{widetext}

The value of average output $\bar{\cal O}_2$ thus saturates at $-\Omega/2\gamma\ll-1$ in the limit of short ${\cal T}\ll \gamma/\Omega^2$. Note that this sudden jump behavior, at a time scale of $\gamma/\Omega^2$  now is only visible at the time-integrated output of the variable ${\cal O}_2$ not commuting with the Hamiltonian. Thus, to achieve a simultaneous sudden jump  for both time-integrated outputs,  we need to require that the Hamiltonian does not commute with both variables .\\
Let us modify the Hamiltonian to $\hat{H}_q=\hbar\frac{\Omega_x}{2}\hat{\sigma}_x+\hbar\frac{\Omega_y}{2}\hat{\sigma}_y$. The joint characteristic function can be written as 

\begin{widetext}

\begin{equation}
\label{cfsmalltimes}
C(\chi, {\cal T}) = \frac{4\gamma  + {\cal T}\left((\Omega_x-ia^{(2,2)}_{VQ}\chi_2)^2-(i\Omega_y-a^{(1,1)}_{VQ}\chi_1)^2\right)}{4\gamma + {\cal T}\bar{\Omega}^2}e^{-\frac{1}{2}\sum_iS^{(i,i)}_{VV}\chi^2_i{\cal T}},
\end{equation}

\end{widetext}
where $\bar{\Omega}^2=\left(\Omega_x^2+\Omega_y^2\right)$. 
This gives  the average outputs
\begin{equation}
\begin{split}
\bar{\cal O}_1 = \frac{2\Omega_y}{4\gamma + {\cal T}\bar{\Omega}^2};
\end{split}
\quad
\begin{split}
\bar{\cal O}_2 = -\frac{2\Omega_x}{4\gamma + {\cal T}\bar{\Omega}^2}.
\end{split}
\end{equation}

Therefore, both time-integrated outputs exhibit a sudden jump at a time scale of $\gamma/\bar{\Omega}^2$. The joint probability distribution of measurement outcomes can then be computed by Fourier transformation of the joint characteristic function \eqref{cfsmalltimes}, and is given by

\begin{widetext}

\begin{equation}
\label{smalltimesxy}
P({\cal O}_1, {\cal O}_2) = \frac{1}{4\gamma + {\cal T}\bar{\Omega}^2}\left(4\gamma+{\cal T}\left[\left(\Omega_x-\frac{4{\cal O}_2}{t_{a2}}\right)^2+\left(\Omega_y+\frac{4{\cal O}_1}{t_{a1}}\right)^2-\frac{4}{{\cal T}t_{a2}}-\frac{4}{{\cal T}t_{a1}}\right]\right)P_G({\cal O}_1, {\cal O}_2).
\end{equation}

\end{widetext}
Here, $t_{ai}\equiv 4S^{(i,i)}_{VV}/\abs{a^{i,i}_{VQ}}^2$ are the acquisition times corresponding to each detector.\\
For the simplest case of identical but independent detectors, $t_{a1}=t_{a2}$, this distribution is positive as long as $K\equiv\gamma t_a\geq 1$ ($K=1$ corresponding to an ideal detector), which is always guaranteed by the corresponding Cauchy-Schwartz inequality \eqref{ineqideal}.

It is instructive to inspect the forms of the distributions \eqref{smalltimes} and \eqref{smalltimesxy} to understand the main characteristics of such a measurement scenario. We do that by plotting the joint distributions and several cross sections of these joint distributions as the distribution of one integrated output given a specific result for the other integrated output. In Figures \ref{fig:smallanalytical} and \ref{fig:smallxyanalytical} we present this two distributions for a measurement time ${\cal T}=4/{\Omega}^2 t_a$ and ${\cal T}=4/\bar{\Omega}^2 t_a$ respectively. The first plot, (a) presents the joint distribution covering a huge range of detector outcomes due to the short  measurement time ${\cal T}$. The sudden jump behavior of the integrated output is visible at this time scale. The position of the peaks and the average integrated outputs in the $({\cal O}_1,{\cal O}_2)$ plane for these distributions depend only on the choice of the Hamiltonian dynamics, as can be seen by comparing Figures \ref{fig:smallanalytical} and \ref{fig:smallxyanalytical}.

In Fig. \ref{fig:smallanalytical} (b) and (c) we present cross sections of the joint distribution \eqref{smalltimes}. First, due to the asymmetry of the Hamiltonian with respect to the two detector outputs, the distributions for ${\cal O}_1$ (plot (a)) are intrinsically different than the distributions for ${\cal O}_2$ (plot (b)). While the average value of the integrated output ${\cal O}_1$ corresponding to the measurement of $\hat{\cal O}_1=\hat{\sigma}_x$ is zero, the average integrated output of the second variable $\hat{\cal O}_2=\hat{\sigma}_y$ can reach anomalously big values as explained by theory of weak values \cite{BigWeakValues}. Figure (b) also shows how conditioning on results of the second integrated output, can be used to drastically change the distribution of the first integrated output, going from a noise-dominated distribution (full line curve for ${\cal O}_2=-20$) to a well-resolved measurement (dashed-dotted curve for ${\cal O}_2=20$).

As noted above, with a proper choice of Hamiltonian, one can achieve anomalously large average integrated outputs in both variables. Thus, now using $\hat{H}_q=\hbar\frac{\Omega_x}{2}\hat{\sigma}_x+\hbar\frac{\Omega_y}{2}\hat{\sigma}_y$, in Fig. \ref{fig:smallxyanalytical} (b) and (c) we present cross sections of the joint distribution \eqref{smalltimesxy}. Here, the asymmetry between the two distributions (a) and (b) disappears and the maximum and minimum values are the same due to our choice of parameters ($\Omega_x=\Omega_y$). 

In Appendix \ref{sec:Positivity}, we use the analytical results for the distribution in the limit of short time and zero overlap to check the positivity of the distribution of measurement outcomes for a more general set of detector noises and responses. We show that the positivity of the distribution is guaranteed provided the restriction \eqref{ineq2}.\\

\section{Numerical results: Short time scales}
\label{sec:NumShort}
In this section, we are going to numerically compute the full probability distribution of measurement outcomes in the same regime as in the previous section, but for experimental conditions. The measurement time is  short compared to the Hamiltonian dynamics of our qubit and the state of the measured system does not vary significantly during this measurement time. For simplicity, in the reminder of this paper we will always consider vanishing cross noises, $S^{(i,i)}_{QV}=0$, for a set of identical but independent detectors. However, the results can be numerically simulated and extended to any two variable measurement scenario.

\begin{figure*}[!ht]
\centering
\includegraphics[width=2.3\columnwidth]{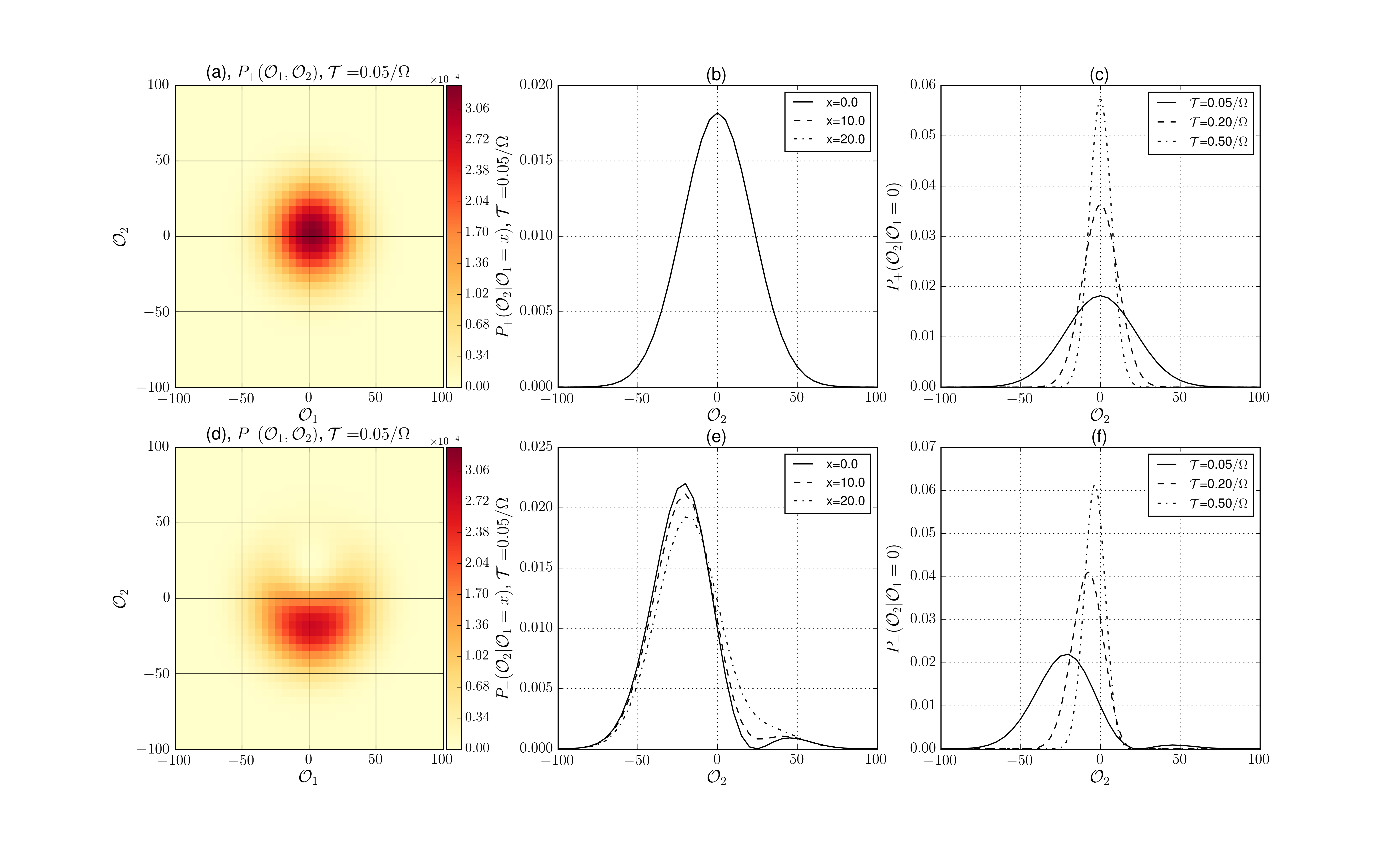}
\caption{(Color online) Output distributions for the simultaneous measurement of $\hat{\cal O}_1=\hat{\sigma}_x$ and $\hat{\cal O}_2=\hat{\sigma}_y$. The measurement with ideal detectors (case (i)) for ${\cal T}\ll \Omega^{-1}$ or comparable with $\Omega^{-1}$. The qubit is prepared in $\ket{Z^{+}}$ and post selected at the end of the measurement: In $\ket{Z^{+}}$ for the first row of figures (plots (a), (b) and (c)); and in $\ket{Z^{-}}$ for the second row of figures (plots (d), (e) and (f)).}
\label{fig:smallideal}
\end{figure*}

\begin{figure*}[!ht]
\centering
\includegraphics[width=2.3\columnwidth]{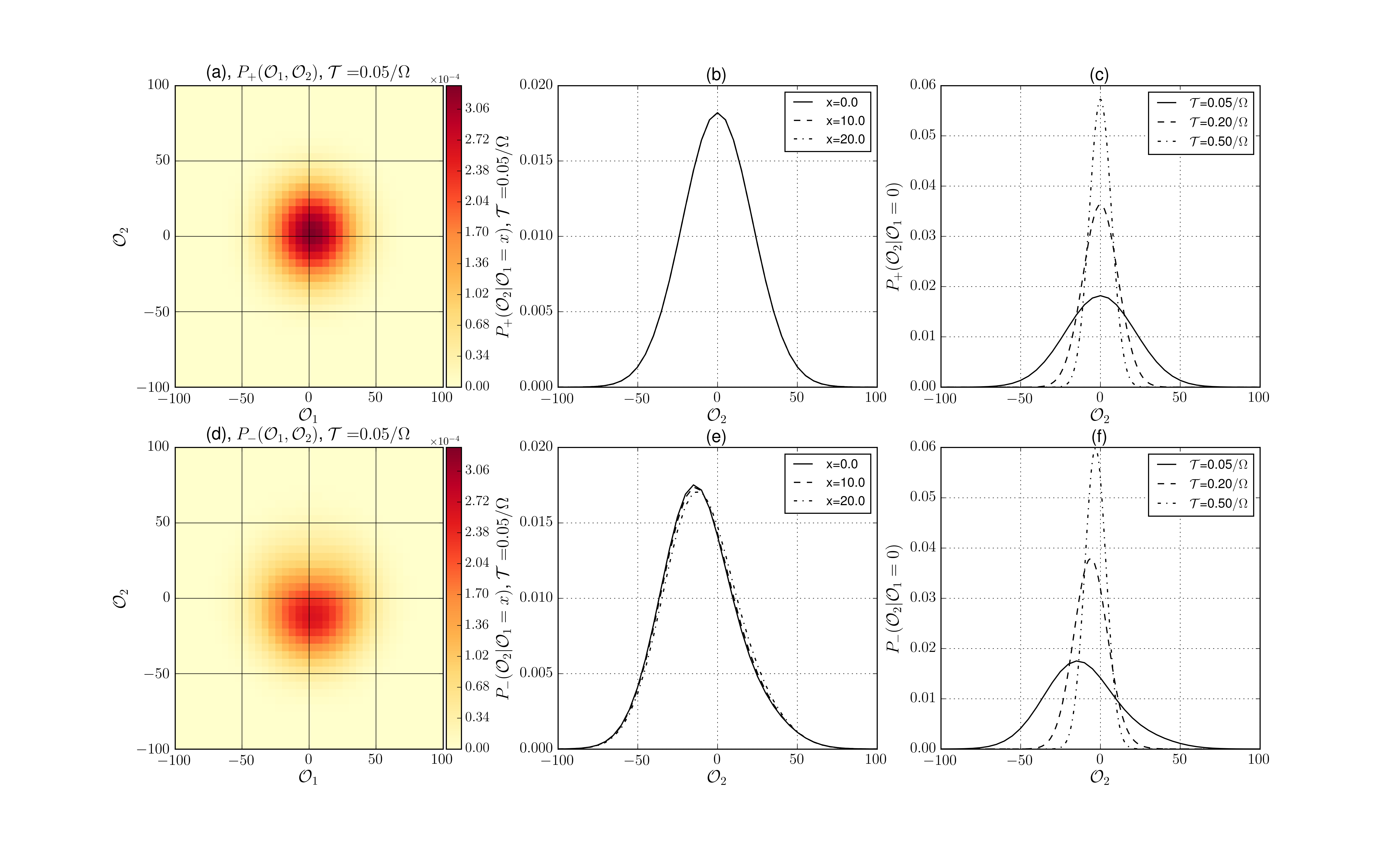}
\caption{(Color online) The measurement with non-ideal detectors and experimentally relevant parameters, case (ii). The qubit is prepared in $\ket{Z^{+}}$ and post selected at the end of the measurement: In  $\ket{Z^{+}}$ state for the first row of figures (plots (a), (b) and (c)); and in $\ket{Z^{-}}$ for the second row of figures (plots (d), (e) and (f)).}
\label{fig:smallexp}
\end{figure*}

\begin{figure*}[!ht]
\centering
\includegraphics[width=2.3\columnwidth]{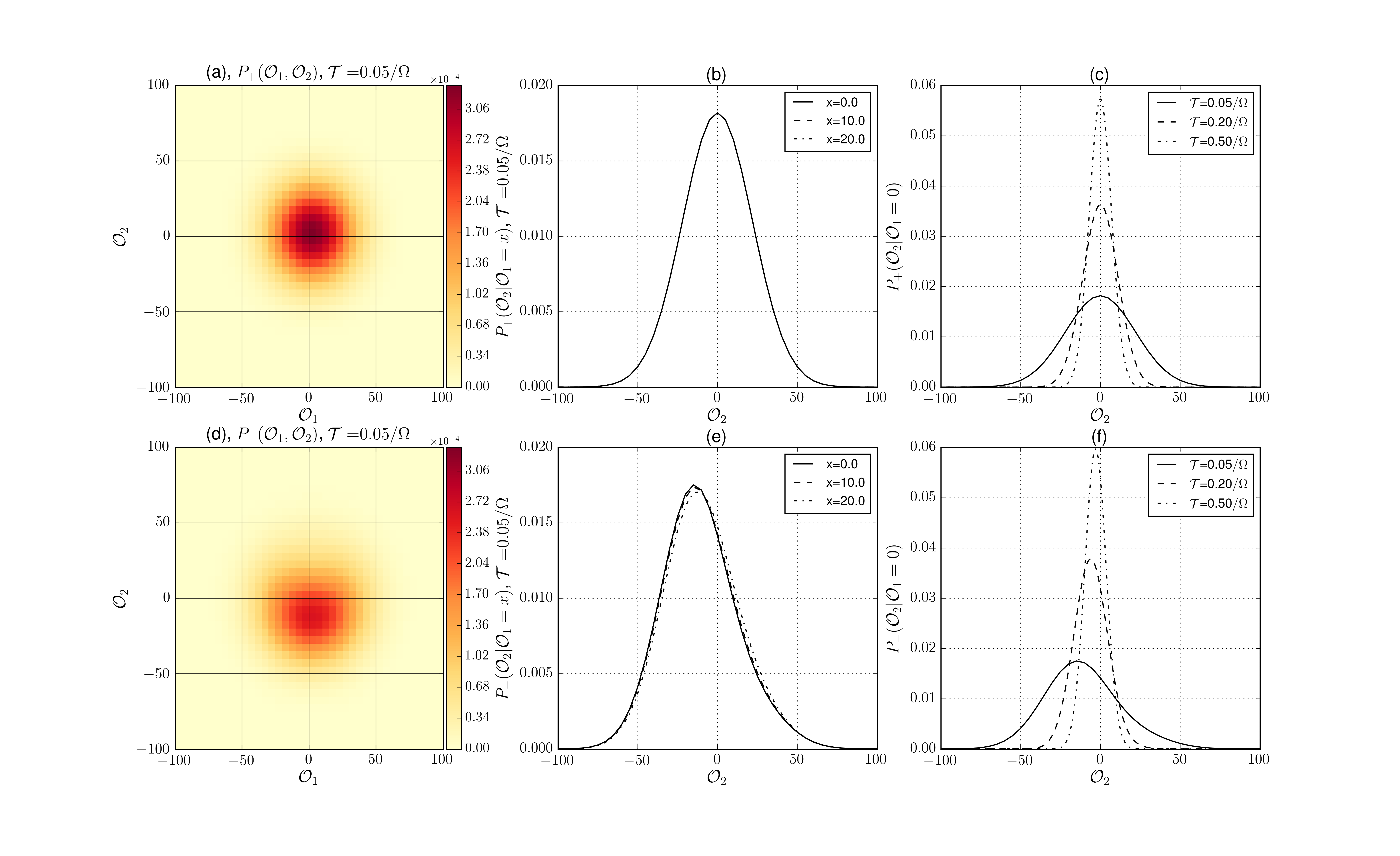}
\caption{(Color online) The measurement with non-ideal detectors and strong detuning, case (iii). The qubit is prepared in $\ket{Z^{+}}$ and post selected at the end of the measurement: In $\ket{Z^{+}}$ for the first row of figures (plots (a), (b) and (c));  and in $\ket{Z^{-}}$ for the second row of figures (plots (d), (e) and (f)).}
\label{fig:smallexpdet}
\end{figure*}

To numerically study this limit and longer time intervals in the next Sec. \ref{sec:NumLong}, let us focus on 3 interesting cases:
\begin{enumerate}[(i)]
\item An ideal detection case, where we numerically solve Eq. \eqref{eq1} with $\hat{H}_q=\hbar\frac{\Omega}{2}\hat{\sigma}_x$ and parameter values such that the inequality \eqref{ineqideal} becomes an equality. Meaning, all the decoherence is brought by the detectors back action and their rates assume the minimum permitted values, $K= t_a\gamma=1$.
\item An experimentally relevant case, where we numerically solve Eq. \eqref{eq1exp} with $\hat{H}_q=\hbar\frac{\Omega}{2}\hat{\sigma}_x$, $\gamma_{\downarrow} = (22.5 \mu\text{s})^{-1}, \gamma_{\uparrow} = (56 \mu\text{s})^{-1}$ and $\gamma_d = (15.6 \mu\text{s})^{-1}$. The acquisition time comes from the measurement rate $2/t_{a}\approx(92 \mu\text{s})^{-1}$ as given in \cite{Huard}.
\item Finally, another experimentally relevant case, where we will again solve Eq. \eqref{eq1exp} but with a modified Hamiltonian, in which a rather strong detuning $\Delta\approx1.7\Omega$ is applied to the qubit as $\hat{H}_q=\hbar\frac{\Omega}{2}\hat{\sigma}_x + \hbar\frac{\Delta}{2}\hat{\sigma}_z$. This value is chosen to maximize $\langle{\cal O}_1\rangle$ for the equilibrium density matrix.
\end{enumerate}

The distributions of the measurements for these three cases are presented in three different figures. Figure \ref{fig:smallideal} for the ideal case, and Figures \ref{fig:smallexp} and \ref{fig:smallexpdet} for the experimentally relevant scenario without and including a strong qubit detuning respectively.

In these three figures, we plot the joint distribution for different combinations of preparation and post-selection states of the measured qubit. As well as cross sections of this distribution, meaning the distribution of a particular detector output given specific values for the other detector output. The first row of plots, (a), (b) and (c), presents these distributions for a qubit prepared in $\ket{Z^{+}}$ and post-selected after the measurement in $\ket{Z^{+}}$; we refer to this as $P_+$. The second row of plots, (d), (e) and (f), presents these distributions for a qubit prepared in $\ket{Z^{+}}$ and post-selected in $\ket{Z^{-}}$; we refer to this as $P_-$.\\
Also, the first column of plots, (a) and (d), are density plots of the joint distribution of measurement outcomes ($P_+({\cal O}_1,{\cal O}_2)$ and $P_-({\cal O}_1,{\cal O}_2)$) for both measured variables and for the measurement time ${\cal T}= 0.05\Omega^{-1}$. The second column of plots, (b) and (e), presents different conditioned distributions of the detector output ${\cal O}_2$ given specific values ${\cal O}_1=x$ of the other detector output ($P_+({\cal O}_2|{\cal O}_1=x)$ and $P_-({\cal O}_2|{\cal O}_1=x)$), again for a measurement time ${\cal T}= 0.05\Omega^{-1}$. Finally, the third column of plots, (c) and (f), presents different conditioned distributions of the detector output ${\cal O}_2$ given a result of ${\cal O}_1=0$ of the other detector output ($P_+({\cal O}_2|{\cal O}_1=0)$ and $P_-({\cal O}_2|{\cal O}_1=0)$) for different measurement times ${\cal T}=0.05,0.2,0.5\Omega^{-1}$.

At this short measurement times, one expects these distributions to be dully Gaussian spreading over a large range of detector output values. This is seen in the upper row of plots. There is only one particular case, as we have shown previously, where this is not true. When the overlap between the preparation and post-selection states is zero. In second row of plots in Figures \ref{fig:smallideal}, \ref{fig:smallexp} and \ref{fig:smallexpdet}, a sudden jump behavior in the averaged integrated output appears, manifested in these figures as very non-Gaussian distribution shapes. There are small deviations in this numerical results because ${\cal T}$ is finite. The plots show anomalously large values for the average integrated output as big shifts in the distribution peaks, in agreement with the analytical results of the previous section. The agreement is visible if one compares Fig. \ref{fig:smallideal} (e) with Fig. \ref{fig:smallanalytical} (c).\\
As expected, this peculiarity is suppressed as the Hamiltonian dynamics start to be relevant (${\cal T}\sim \Omega^{-1}$) as can be seen in the different curves at increasing time intervals in the third column of plots in Figures \ref{fig:smallideal}, \ref{fig:smallexp} and \ref{fig:smallexpdet}. The shape of the distributions becomes more Gaussian as the detectors considered are less ideal. This can be seen when comparing the distributions for ideal detectors (Fig. \ref{fig:smallideal}) and experimentally relevant parameters (Fig. \ref{fig:smallexp} and \ref{fig:smallexpdet}).\\
As the measurement time is short compared to the Hamiltonian dynamics, the qubit state changes insignificantly during the measurement. This fact is manifested in the sudden jump behavior of the $P_-$ distributions in the second row of plots, and in the fact that Figures \ref{fig:smallexp} and \ref{fig:smallexpdet} are almost the same. At these short measurement times, a difference in the Hamiltonian is not noticeable.

\section{Numerical results: Longer time scales}
\label{sec:NumLong}
In the previous section, we have presented the distributions of CWLM outcomes of the simultaneous measurement of two non-commuting variables in the limit of short measurement times. In this Section, we address the distributions of the CWLM outcomes of the simultaneous measurement of two non-commuting variables at time scales of the order of coherence/relaxation times and $t_a$.\\

\begin{figure*}[!ht]
\centering
\includegraphics[width=2.3\columnwidth]{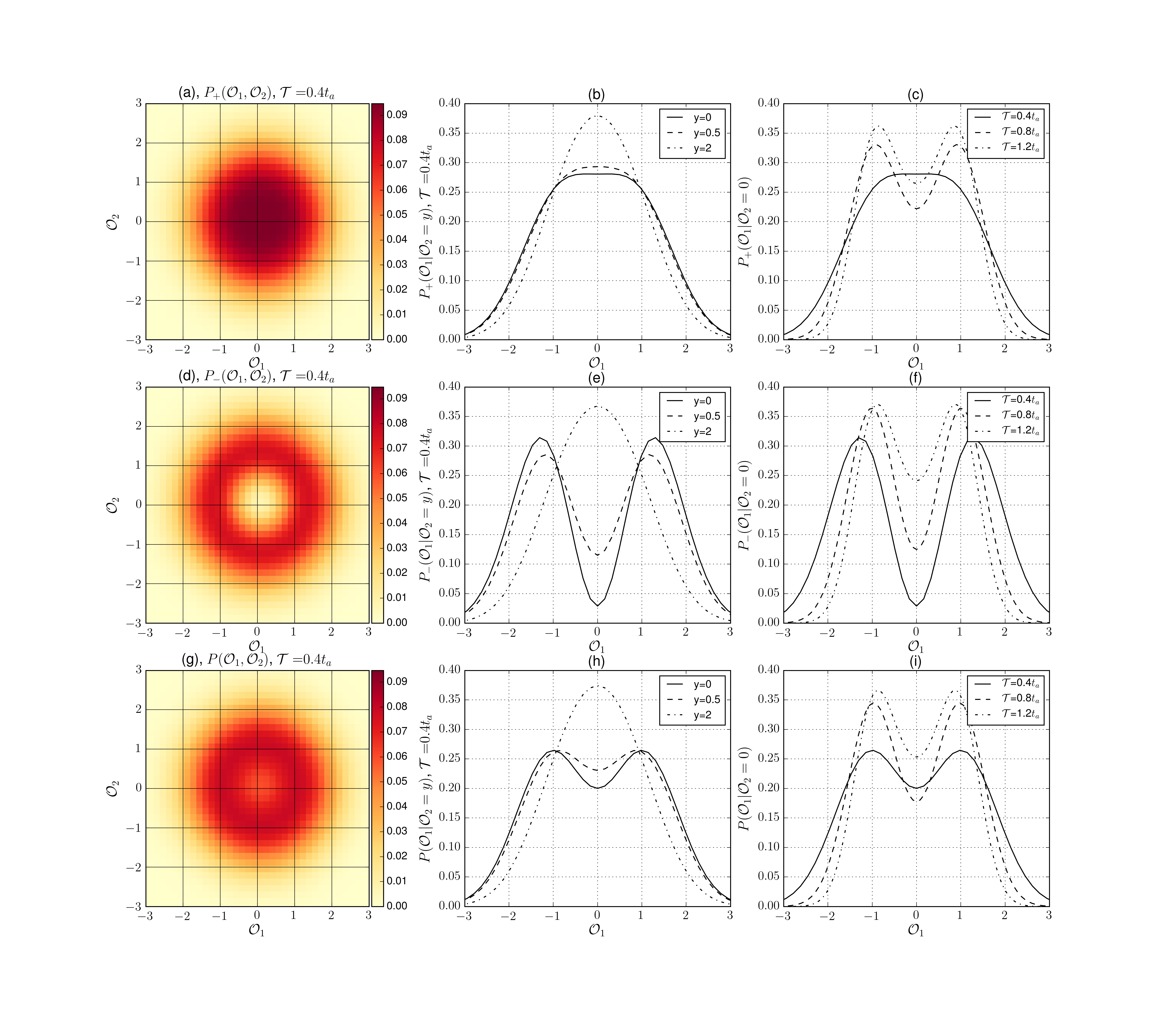}
\caption{(Color online) Output distributions for the simultaneous measurement of $\hat{\cal O}_1=\hat{\sigma}_x$ and $\hat{\cal O}_2=\hat{\sigma}_y$. The qubit is prepared in $\ket{Z^{+}}$ and post selected either in $\ket{Z^{+}}$ for the first row of figures (plots (a), (b) and (c)); or in $\ket{Z^{-}}$ for the second row of figures (plots (d), (e) and (f)). There is no post-selection for the last row of figures (plots (g), (h) and (i)). The measurement is performed with ideal detectors and no Hamiltonian dynamics are present during time intervals comparable to the acquisition time of the measurement setup.\\
In this configuration, the peculiarities discussed in Sec. \ref{sec:quasidistributions} are clearly visible in the joint distributions (plots (a), (d) and (g)).}
\label{fig:idealnodrive}
\end{figure*}

\begin{figure*}[!ht]
\centering
\includegraphics[width=2.3\columnwidth]{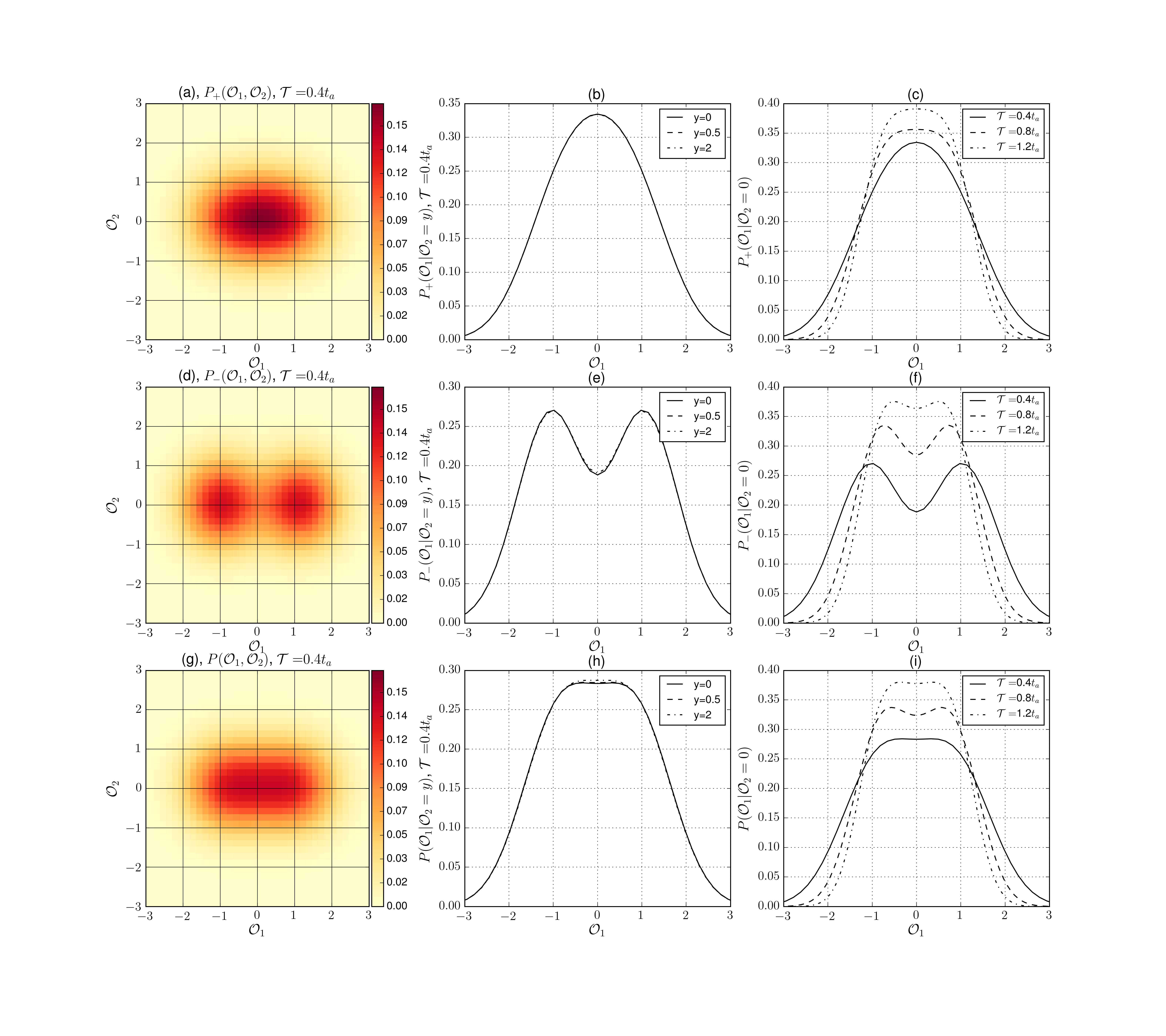}
\caption{(Color online)Here, the output distributions for a measurement performed with ideal detectors (case (i)) during time intervals comparable to the acquisition time of the measurement setup. The Hamiltonian drastically changes the type of peculiarities seen in the distributions. This can be seen by comparison with Figure \ref{fig:idealnodrive}.}
\label{fig:ideal}
\end{figure*}

To begin with, let us assume no Hamiltonian and ideal detectors (case (i) with no Hamiltonian). With this, the conditioned distribution of outcomes changes only at the time scale $t_a \simeq \gamma^{-1}$, that is much longer than $\Omega^{-1}$, and the dynamics are described by Eq. \eqref{eq1} with vanishing  $S^{(i,i)}_{QV}$ and $\hat{H}_q$ terms.\\
As discussed in Section \ref{sec:quasidistributions}, the pre and post-selection condition leads to peculiarities in the joint distribution that are located around the circle ${\cal O}_1^2+{\cal O}_2^2=1$. These peculiarities should be visible in these conditions, at intermediate measurement times that are longer than $\Omega^{-1}$ but short enough so that they are comparable to $t_a\simeq \gamma^{-1}$.\\
In fact, this is what we present in Fig. \ref{fig:idealnodrive} when plotting the joint distributions (first column of plots; (a), (d) and (g)). The cross sections ((b), (e) and (h); and (c), (f) and (i)) show the expected half-quantization peculiarities characteristic of a single variable measurement, extensively discussed in \cite{FranquetNaz}.
In the last column of plots ((c), (f) and (i)) one could expect the separation of the distribution onto peaks at the limit ${\cal T}\gg t_a$. However, this is a signature of a quantum nondemolition measurement, and the fact that we inspect the simultaneous measurement of two non-commuting variables means that the measurement itself will induce rates causing transitions between the qubit states. Thus, not being a nondemolition measurement.\\
It is worth mentioning that the fact that all these distributions are symmetric under a change  ${\cal O}_1\leftrightarrow{\cal O}_2$ is due to the choice identical detectors and no Hamiltonian dynamics in any of those variables axes ($\hat{\cal O}_1=\hat{\sigma}_x$ and $\hat{\cal O}_2=\hat{\sigma}_y$).

To clarify these observations, let us describe precisely the layout both for Fig. \ref{fig:idealnodrive} and Fig. \ref{fig:ideal}: The (a) plots show the joint distribution of measurement outcomes $P_+({\cal O}_1,{\cal O}_2)$ for a qubit prepared in $\ket{Z^{+}}$ and post-selected in the same state after the measurement of duration ${\cal T}=0.4t_a$. The (d) plots show the joint distribution of measurement outcomes $P_-({\cal O}_1,{\cal O}_2)$ for a qubit prepared in $\ket{Z^{+}}$ and post-selected in the orthogonal state $\ket{Z^{-}}$ after the measurement of duration ${\cal T}=0.4t_a$. The (g) plots show the joint distribution of measurement outcomes $P({\cal O}_1,{\cal O}_2)$ for a qubit prepared in $\ket{Z^{+}}$ unconditioned to any post-selection after the measurement of duration ${\cal T}=0.4t_a$. Next, in the second column, the (b) plots present the conditioned distributions $P_+({\cal O}_1|{\cal O}_2=y)$ of the first output, given a result ${\cal O}_2=y$ for the second output, again for a qubit prepared in $\ket{Z^{+}}$ and post-selected in the same state after the measurement of duration ${\cal T}=0.4t_a$. Respectively, the (e) and (h) plots, show the conditioned distributions $P_-({\cal O}_1|{\cal O}_2=y)$ and $P({\cal O}_1|{\cal O}_2=y)$. Finally, in the third column, we plot the conditioned distributions $P_+({\cal O}_1|{\cal O}_2=0)$ in (c), $P_-({\cal O}_1|{\cal O}_2=0)$ in (f), and $P({\cal O}_1|{\cal O}_2=0)$ in (i); for different measurement duration ${\cal T}=0.4,0.8,1.2t_a$.\\
In contrast with the figures in the previous section, as the measurement time is big enough so that the qubit state changes appreciably during the measurement, we also plot the unconditioned distributions $P({\cal O}_1,{\cal O}_2)$ now being clearly different than the distributions conditioned to a specific pot-selection $P_{\pm}({\cal O}_1,{\cal O}_2)$.\\

Let us incorporate Hamiltonian dynamics to this measurement scenario, focusing now on case (i). If we keep the final state fixed to $\ket{Z^{\pm}}$, the contribution due to the conditioned evolution in these distributions will exhibit fast oscillations as function of ${\cal T}$ with a period $2\pi/\Omega$. It is proficient from both theoretical and experimental considerations to quench these rather trivial oscillations. We achieve this by projecting the qubit after the measurement on the states  $|\bar{Z}^{\pm}\rangle = e^{-i \hat{H}_q {\cal T}} |Z^{\pm} \rangle$ thereby correcting for the trivial qubit dynamics. In practice, such correction can be achieved by applying a short pulse rotating the qubit about $x$-axis right before the post-selection measurement.\\
With this, the asymmetry in the Hamiltonian with respect to the measured $\hat{\cal O}_1$ and $\hat{\cal O}_2$ variables will break the symmetry in the shape of the distributions. Then, the conditioned distributions for the output ${\cal O}_2$ are just Gaussian functions centered at ${\cal O}_2=0$ with their spread decreasing over time as $\sim 1/ \sqrt{\cal{T}}$. Thus giving no information about the output ${\cal O}_2$ at this time scale. That is why we choose to plot the conditioned distributions for the output ${\cal O}_1$.
This situation is presented in Fig. \ref{fig:ideal}. The choice of Hamiltonian now collapses all these peculiarities due to the pre and post-selection conditions in one of the two outputs. This is perfectly visible in the shape of the joint distributions (plots (a), (d) and (g)). Not only that, but the addition of dynamics to the measured qubit results in a clear and strong suppression of the dependence of a given output on the other output outcomes as can be seen in plots (b), (e) and (h). Finally, if one compares the time evolution of these distributions (plots (c), (f) and (i)) for figures \ref{fig:idealnodrive} and \ref{fig:ideal}, the addition of dynamics to the measured qubit, results in a less resolved measurement, i.e., less separated peaks for a given measurement time $\mathcal{T}$.

Although this shows that the interference effect and peculiarities due to conditioned evolution are still visible at longer time scales for an ideal measurement scenario, it is also clear, that those signatures are suppressed by dynamics in the measured qubit. In fact, in an experimental situation, where external sources of decoherence are present, resolving those signatures might become a very challenging task. It is then important to inspect an experimentally relevant parameter regime in these numerical simulations.

\begin{figure*}[!ht]
\centering
\includegraphics[width=1.5\columnwidth]{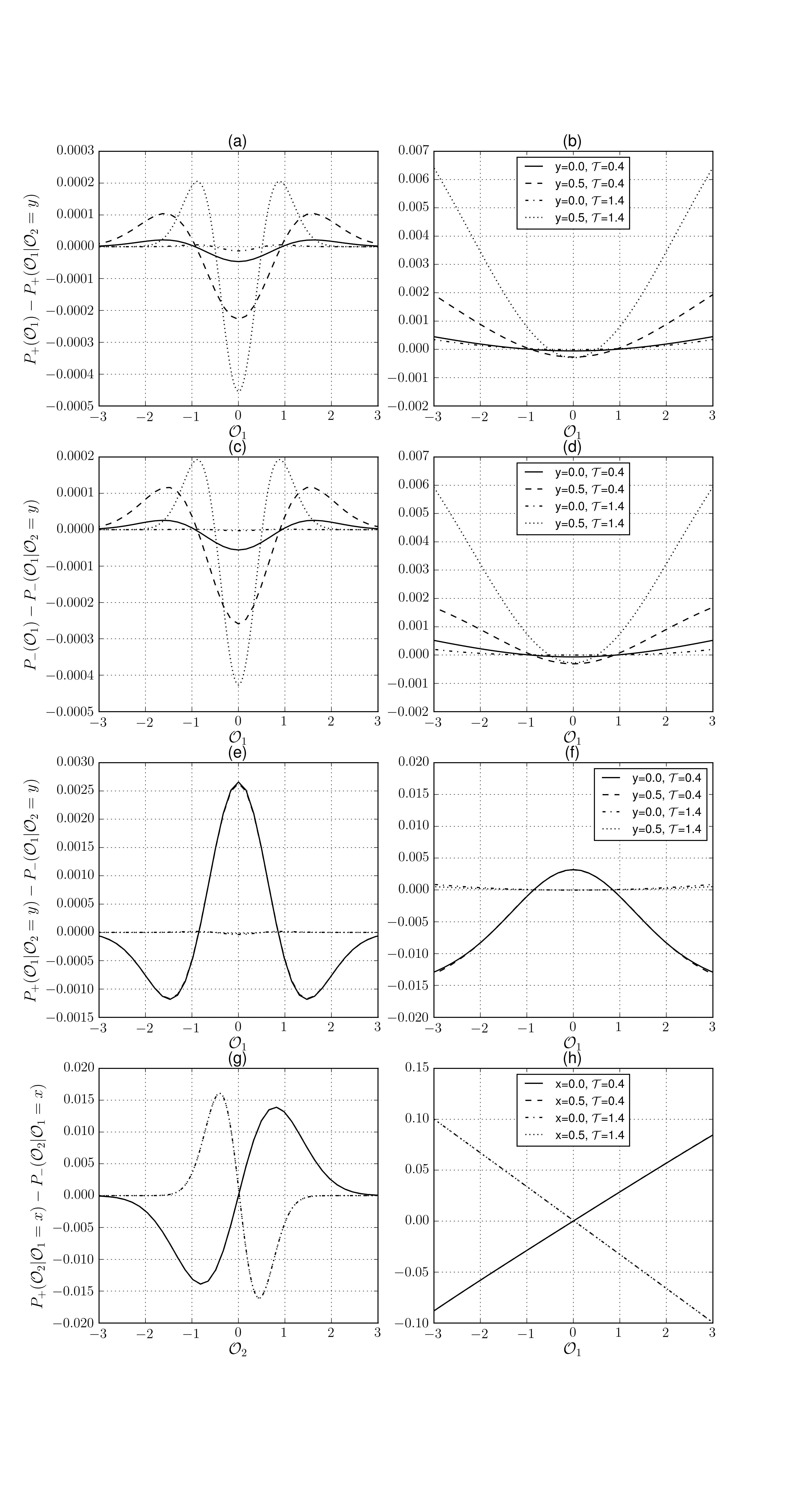}
\vspace*{-25mm}
\caption{Differences of distributions at measurement times of the order of the acquisition time $t_a$ under experimentally relevant measurement conditions, case (ii). In the first column of plots different distributions are examined to understand the correlations between outputs and time evolution in (a) and (c), as well as the visibility of the conditioned evolution peculiarities in the statistics of both outputs in (e) and (f). In the second column, the plotted difference of probabilities is divided by the sum of the same probabilities. The layout is described in detail in Sec. \ref{sec:NumLong} in the main text.}
\label{fig:expdiff}
\end{figure*}

\begin{figure*}[!ht]
\centering
\includegraphics[width=1.5\columnwidth]{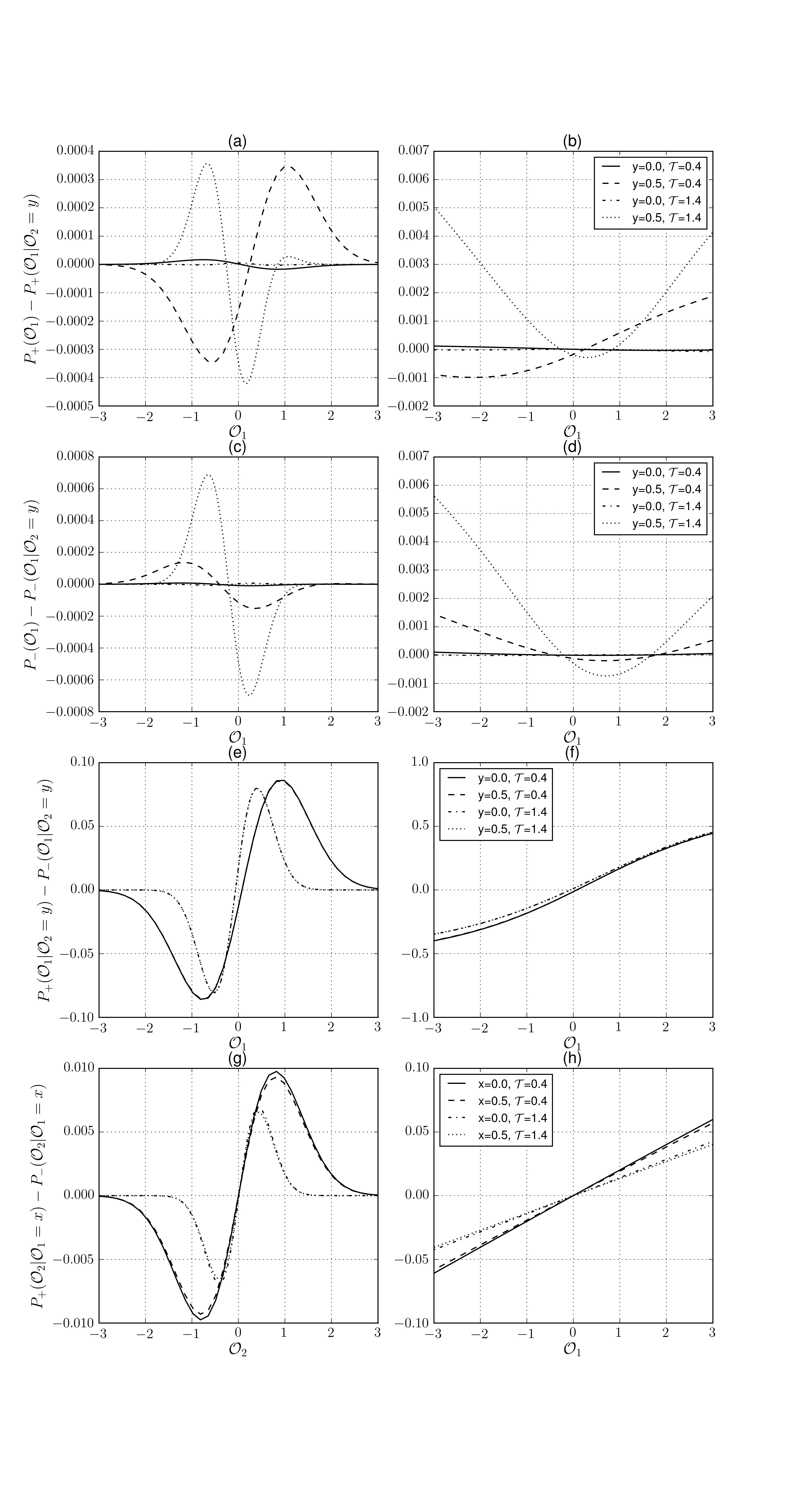}
\vspace*{-25mm}
\caption{Differences of distributions taken at measurement times of the order of the acquisition time $t_a$ and at nonzero detuning, case (iii). The layout is the same as in previous Figure \ref{fig:expdiff}.}
\label{fig:expdetdiff}
\end{figure*}

To this end, one can inspect experimentally relevant scenarios like cases (ii) and (iii). It is good to note that the quality of the measurement setup in these conditions is far from ideal, $K = t_a \gamma_d \approx 12$, and at longer time scales, the decoherence completely dominates all the measurement dynamics. It is so that the distributions do not show visible characteristics of the conditioned qubit evolution. They appear to be just Gaussian distributions centered at zero value of the outcome variables.

In this case, it is more instructive to inspect the difference of two particular distributions, rather than the distribution itself. With that in mind, in Figures \ref{fig:expdiff} and \ref{fig:expdetdiff} we plot different differences of distributions. In Fig. \ref{fig:expdiff} we consider case (ii). In Fig. \ref{fig:expdetdiff} we consider case (iii). In doing so, not only we are interested in the phenomena related to conditioned qubit evolution, but also in the difference of simultaneous measurement of several variables from the single variable case.\\
These two figures are structured with the following layout: The (a) plots, show the difference of the distribution of the first output disregarding the second output and the distribution of the same first output given a specific result $y$ for the second output, $P_+({\cal O}_1) - P_+({\cal O}_1|{\cal O}_2=y)$. The (b) plots, show the same difference divided by its sum, $(P_+({\cal O}_1) - P_+({\cal O}_1|{\cal O}_2=y))/(P_+({\cal O}_1) + P_+({\cal O}_1|{\cal O}_2=y))$. Both for a qubit prepared in $\ket{Z^{+}}$ state and post-selected in the same state. The (c) plots, show again that difference but for a qubit prepared in $\ket{Z^{+}}$ and post-selected in $\ket{Z^{-}}$, i.e., $P_-({\cal O}_1) - P_-({\cal O}_1|{\cal O}_2=y)$. Respectively, (d) show that difference divided by the sum, $(P_-({\cal O}_1) - P_-|({\cal O}_1|{\cal O}_2=y))/(P_-({\cal O}_1) + P_-({\cal O}_1|{\cal O}_2=y))$. These differences give an estimation of the correlation between the two outputs in these measurements, or the separability of the joint distribution.\\
Next, the (e) plots show the difference between the distribution of the first output given a specific result $y$ for the second output for a qubit prepared in $\ket{Z^{+}}$ and post-selected in the same state; and the distribution of the first output given a specific result $y$ for the second output for a qubit prepared in $\ket{Z^{+}}$ and post-selected in $\ket{Z^{-}}$. That is $P_+({\cal O}_1|{\cal O}_2=y)-P_-({\cal O}_1|{\cal O}_2=y)$. Again, the (f) plots show this difference divided by their sum, $(P_+({\cal O}_1|{\cal O}_2=y)-P_-({\cal O}_1|{\cal O}_2=y))/(P_+({\cal O}_1|{\cal O}_2=y)+P_-({\cal O}_1|{\cal O}_2=y))$. Finally, the (g) plots show the same difference, but for the distributions of the second output given a specific result $x$ for the first output: $P_+({\cal O}_2|{\cal O}_1=x)-P_-({\cal O}_2|{\cal O}_1=x)$. Respectively (h) show that difference divided by their sum, $(P_+({\cal O}_2|{\cal O}_1=x)-P_-({\cal O}_2|{\cal O}_1=x))/(P_+({\cal O}_2|{\cal O}_1=x)+P_-({\cal O}_2|{\cal O}_1=x))$.\\
The reason for inspecting these last differences is simple, we want to have an estimation for the resolution of any signature that is due to the conditioned evolution of the measured system. Thus, inspecting the difference between the two limiting cases of this conditioned evolution, i.e., same pre and post-selection and orthogonal pre and post-selection; shows how strong these signatures are. Furthermore, these differences divided by their sums, quantify how much these signatures can be resolved by using the output distributions of such measurements. Or in other words, the certainty with which one can distinguish two distributions from each other given a measurement reading:\cite{FranquetNaz}
\begin{equation}
C({\cal O}_i|{\cal O}_j=\alpha) =\frac{P_+({\cal O}_i|{\cal O}_j=\alpha)-P_-({\cal O}_i|{\cal O}_j=\alpha)}{P_+({\cal O}_i|{\cal O}_j=\alpha)+P_-({\cal O}_i|{\cal O}_j=\alpha)}
\end{equation}

The values $C = {\pm}1$ would imply that the measurement is {\it certainly} post-selected in $\ket{Z^{\pm}}$.\\

In this experimental regime at zero detuning, Fig. \ref{fig:expdiff}, the differences of distributions (a) and (c), reveal that the two outputs are still correlated, and this correlation seems to be bigger for given values of the outputs that are farther away from the origin where the distributions peak at such time scales. Nevertheless, it is very small, as the joint distribution quickly becomes a Gaussian due to decoherence and relaxation. At big values of ${\cal O}_1$, the difference quickly decreases together with the distributions. In this respect, it is instructive to inspect the difference normalized on the sum of the probability densities. As we see from (b) and (d), this quantity increases with increasing ${\cal O}_1$, reaches relatively large values at increasing ${\cal O}_2=y$ results due to their low statistical weight, and seems to remain relevant at a small region ${\cal O}_1\sim 0$ even for big times. This region is more relevant because this quantity is not suppressed or increased due to exponentially low probabilities for those values, it is rather a direct measure of the correlation of the two outputs.\\
The signatures of the conditioned evolution are revealed by the differences in (e) and (g). As expected due to the form of the Hamiltonian (on $\hat{\cal O}_1=\hat{\sigma}_x$ axis), (e) is very different from (g). In (e), the shape of the difference suggests that the $P_-({\cal O}_1|{\cal O}_2=y)$ is pushed on both positive and negative values of ${\cal O}_1$ in comparison with $P_+({\cal O}_1|{\cal O}_2=y)$, in agreement with the previous findings. The decoherence and relaxation quickly diminish the difference upon increasing ${\cal T}$. Inspecting the certainty in (f), it saturates with increasing ${\cal O}_1$, reaches relatively large values at short ${\cal T}$, and fades away upon increasing ${\cal T}$. Note that at short ${\cal T}=0.4t_a$ this relative difference achieves $0.002$ at ${\cal O}_1\approx 0$ and can be thus revealed from the statistics of several hundred individual measurements.
For the second output ${\cal O}_2$, the differences in (g) are an order of magnitude bigger than those for the ${\cal O}_1$ distributions in plot (e). Not only that, but the difference does not vanish in the limit of big ${\cal T}$. Rather, it is concentrated in an increasingly narrow interval of ${\cal O}_2$ conforming to the decreasing width of the distribution. It is worth noting that they also change sign as ${\cal T}$ increases. For short times, the shape of $P_-({\cal O}_2|{\cal O}_1=x)$ resembles the shape of the distribution at the sudden jump regime (Fig. \ref{fig:smallexp}), as the time ${\cal T}$ increases, the distributions are shifted depending on the post-selected state. This is not the case for the ${\cal O}_1$ output discussed previously because of the chosen Hamiltonian. As to the certainty in (h), it shows a linear behavior with ${\cal O}_2$, $C({\cal O}_2|{\cal O}_1=x)=\beta {\cal O}_2$. The sign of $\beta$ depends on the sign of the shift in the distributions, and the linear behavior can be explained in the limit of small shifts. This does not imply that the distributions are different in this limit since they become concentrated with divergent probability density, and the values of ${\cal O}_2$ with high certainty occur with exponentially low probability. This is discussed in detail Section V of\cite{FranquetNaz}.\\

Let us inspect these differences of distributions at nonzero detuning in Fig. \ref{fig:expdetdiff}. In this case, there is no reason to expect the ${\cal O}_1\rightarrow -{\cal O}_1$ symmetry in the distributions, or in turn, the differences. Again the differences showed in (a), (b), (c) and (d); reveal small correlations between the two outputs still in the presence of detuning. These are bigger when the distributions are conditioned with values at bigger distances from the origin. 
The differences of the probability distributions presented in (e) and (g) seem to be at least one order of magnitude bigger for the distributions of the first output in (e), compared to the zero detuning case in Fig. \ref{fig:expdiff}. And, for both outputs, the difference does not vanish in the limit of big ${\cal T}$ . Rather, it is concentrated in an increasingly narrow interval of ${\cal O}_{1,2}$ conforming to the decreasing width of the distribution.
This suggests that adding a strong detuning can increase the detection resolution, and reveal these distribution differences from the statistics of fewer individual measurements. However, as mentioned before, the certainties (f) and (g), rather quickly converge when increasing ${\cal T}$ to finite and and rather big values in a wide interval of the output ${\cal O}_{1,2}$ in question. Again this do not imply that the distributions are fundamentally different in this limit since they become concentrated with divergent probability densities. For the certainty of the second output distributions in (h), a linear behavior appears due to the small shifts limit of the distributions.\\

It is worth noting, that although the joint distribution of measurement outcomes effectively becomes a product distribution $P({\cal O}_1,{\cal O}_2)\approx P({\cal O}_1)P({\cal O}_2)$, meaning the correlations between the two non-commuting variables are lost rather fast, when increasing ${\cal T}$. The signature of interference due to the conditioned dynamics in the probability distribution can still be revealed from the statistics of several hundred individual measurements in experimental conditions.

\section{Conclusion}
\label{sec:conclusion}

In this work, we put forward a proper theoretical formalism based on full counting statistics approach \cite{NazKin, NazWei} to describe and evaluate the measurement statistics in the course of conditioned quantum evolution. We extend the previous work\cite{FranquetNaz} to the simultaneous measurement of two non-commuting variables. We illustrate this formalism with several examples and prove that the interesting features arising from the conditioned quantum evolution can be seen in the statistics of the measurement outcomes for both short and relatively long measurement intervals. We also reveal the interplay between the two non-commuting variables statistics and the signatures of the conditioned dynamics in the individual and joint distributions.\\

We describe and investigate two signatures of the conditioned statistics that are related to quantum interference effects. First is the appearance of peculiarities at the circle ${\cal O}^2_1+{\cal O}^2_2=1$ in the distribution of measurement outcomes, that is revealed by a quasi-distribution of shifts located at the compact support  ${\cal O}^2_1+{\cal O}^2_2=1$ ,  ${\cal O}^2_1+{\cal O}^2_2=0$ as well as inside the circle  ${\cal O}^2_1+{\cal O}^2_2<1$. This provides a connection with what we termed {\it half-quantized} measurement values for the single variable measurement case, as the distribution function may display peculiarities, that are either peaks or dips, at half-sums of the quantized values.
In the special case of zero overlap between initial and final states and time intervals that are so short as the wave function of the system does not significantly change by either Hamiltonian or dissipative dynamics. We reveal unexpectedly large values of the time-integrated output cumulants for such short intervals, that we termed {\it sudden jump}. We show that a simultaneous jump in integrated output can be achieved in both measured variables given an appropriate choice of Hamiltonian. This effect is felt in a short time scale $\gamma/\Omega^2$ where $\gamma^{-1}$ is the time scale of dissipative dynamics and $\Omega^{-1}$ is the time scale of Hamiltonian dynamics. Additionally our results show that it is possible to achieve bigger saturation values for these anomalously big averages when further conditioning the statistics of one output with the results of other outputs.\\

Our results show that it is possible to have very detailed theoretical predictions of CWLM distributions. In particular, we show how to use this formalism to account for conditioned quantum evolution and simultaneous non-commuting variable measurements in the paradigm of CWLM. This opens the possibility for investigation and characterization of quantum effects even if the choice of parameters is far from ideal and the effects are small.\\

The signatures in the distributions that we predict in this paper can be seen in realistic experimental regimes. One of the key elements to experimentally observe this effects is the ability to efficiently record time traces for a weak continuous monitoring of one or rather, several qubit variables, and this has been achieved in several papers \cite{Devoret, SiddiqiSingle, SiddiqiEntanglement,Huard, DiCarlo, SiddiqiMapping, SiddiqiMolmer, SiddiqiDynamics} applying it for the observation of qubit trajectories or real-time feedback. Thus, we believe it is possible to extract these kind of statistics from the existing records of several experiments.\\

This work was supported by the Netherlands Organization for Scientific Research (NWO/OCW), as part of the Frontiers of Nanoscience program.

\bibliography{paper2d}

\providecommand{\noopsort}[1]{}\providecommand{\singleletter}[1]{#1}%
\begin{thebibliography}{29}%
\makeatletter
\providecommand \@ifxundefined [1]{%
 \@ifx{#1\undefined}
}%
\providecommand \@ifnum [1]{%
 \ifnum #1\expandafter \@firstoftwo
 \else \expandafter \@secondoftwo
 \fi
}%
\providecommand \@ifx [1]{%
 \ifx #1\expandafter \@firstoftwo
 \else \expandafter \@secondoftwo
 \fi
}%
\providecommand \natexlab [1]{#1}%
\providecommand \enquote  [1]{``#1''}%
\providecommand \bibnamefont  [1]{#1}%
\providecommand \bibfnamefont [1]{#1}%
\providecommand \citenamefont [1]{#1}%
\providecommand \href@noop [0]{\@secondoftwo}%
\providecommand \href [0]{\begingroup \@sanitize@url \@href}%
\providecommand \@href[1]{\@@startlink{#1}\@@href}%
\providecommand \@@href[1]{\endgroup#1\@@endlink}%
\providecommand \@sanitize@url [0]{\catcode `\\12\catcode `\$12\catcode
  `\&12\catcode `\#12\catcode `\^12\catcode `\_12\catcode `\%12\relax}%
\providecommand \@@startlink[1]{}%
\providecommand \@@endlink[0]{}%
\providecommand \url  [0]{\begingroup\@sanitize@url \@url }%
\providecommand \@url [1]{\endgroup\@href {#1}{\urlprefix }}%
\providecommand \urlprefix  [0]{URL }%
\providecommand \Eprint [0]{\href }%
\providecommand \doibase [0]{http://dx.doi.org/}%
\providecommand \selectlanguage [0]{\@gobble}%
\providecommand \bibinfo  [0]{\@secondoftwo}%
\providecommand \bibfield  [0]{\@secondoftwo}%
\providecommand \translation [1]{[#1]}%
\providecommand \BibitemOpen [0]{}%
\providecommand \bibitemStop [0]{}%
\providecommand \bibitemNoStop [0]{.\EOS\space}%
\providecommand \EOS [0]{\spacefactor3000\relax}%
\providecommand \BibitemShut  [1]{\csname bibitem#1\endcsname}%
\let\auto@bib@innerbib\@empty
\bibitem [{\citenamefont {Nielsen}\ and\ \citenamefont
  {Chuang}(2000)}]{NielsenChuang}%
  \BibitemOpen
  \bibfield  {author} {\bibinfo {author} {\bibfnamefont {M.}~\bibnamefont
  {Nielsen}}\ and\ \bibinfo {author} {\bibfnamefont {I.}~\bibnamefont
  {Chuang}},\ }\href@noop {} {\emph {\bibinfo {title} {Quantum Computation and
  Quantum Information}}}\ (\bibinfo  {publisher} {Cambridge University Press},\
  \bibinfo {year} {2000})\BibitemShut {NoStop}%
\bibitem [{\citenamefont {Mensky}(1991)}]{CWLM0}%
  \BibitemOpen
  \bibfield  {author} {\bibinfo {author} {\bibfnamefont {M.~B.}\ \bibnamefont
  {Mensky}},\ }\href {\doibase http://dx.doi.org/10.1016/0375-9601(91)90474-M}
  {\bibfield  {journal} {\bibinfo  {journal} {Physics Letters A}\ }\textbf
  {\bibinfo {volume} {155}},\ \bibinfo {pages} {229 } (\bibinfo {year}
  {1991})}\BibitemShut {NoStop}%
\bibitem [{\citenamefont {Korotkov}(1999)}]{CWLM1}%
  \BibitemOpen
  \bibfield  {author} {\bibinfo {author} {\bibfnamefont {A.~N.}\ \bibnamefont
  {Korotkov}},\ }\href@noop {} {\bibfield  {journal} {\bibinfo  {journal}
  {Phys. Rev. B}\ }\textbf {\bibinfo {volume} {60}},\ \bibinfo {pages} {5737}
  (\bibinfo {year} {1999})}\BibitemShut {NoStop}%
\bibitem [{\citenamefont {Jordan}\ and\ \citenamefont
  {Buttiker}(2005)}]{CWLM2}%
  \BibitemOpen
  \bibfield  {author} {\bibinfo {author} {\bibfnamefont {A.}~\bibnamefont
  {Jordan}}\ and\ \bibinfo {author} {\bibfnamefont {M.}~\bibnamefont
  {Buttiker}},\ }\href@noop {} {\bibfield  {journal} {\bibinfo  {journal}
  {Phys. Rev. Lett.}\ }\textbf {\bibinfo {volume} {95}},\ \bibinfo {pages}
  {220401} (\bibinfo {year} {2005})}\BibitemShut {NoStop}%
\bibitem [{\citenamefont {Jacobs}\ and\ \citenamefont {Steck}(2006)}]{CWLM25}%
  \BibitemOpen
  \bibfield  {author} {\bibinfo {author} {\bibfnamefont {K.}~\bibnamefont
  {Jacobs}}\ and\ \bibinfo {author} {\bibfnamefont {D.~A.}\ \bibnamefont
  {Steck}},\ }\href {\doibase 10.1080/00107510601101934} {\bibfield  {journal}
  {\bibinfo  {journal} {Contemporary Physics}\ }\textbf {\bibinfo {volume}
  {47}},\ \bibinfo {pages} {279} (\bibinfo {year} {2006})},\ \Eprint
  {http://arxiv.org/abs/http://dx.doi.org/10.1080/00107510601101934}
  {http://dx.doi.org/10.1080/00107510601101934} \BibitemShut {NoStop}%
\bibitem [{\citenamefont {Wei}\ and\ \citenamefont {Nazarov}(2008)}]{NazWei}%
  \BibitemOpen
  \bibfield  {author} {\bibinfo {author} {\bibfnamefont {H.}~\bibnamefont
  {Wei}}\ and\ \bibinfo {author} {\bibfnamefont {Y.~V.}\ \bibnamefont
  {Nazarov}},\ }\href {\doibase 10.1103/PhysRevB.78.045308} {\bibfield
  {journal} {\bibinfo  {journal} {Phys. Rev. B}\ }\textbf {\bibinfo {volume}
  {78}},\ \bibinfo {pages} {045308} (\bibinfo {year} {2008})}\BibitemShut
  {NoStop}%
\bibitem [{\citenamefont {Chantasri}\ and\ \citenamefont
  {Jordan}(2015)}]{CWLM3}%
  \BibitemOpen
  \bibfield  {author} {\bibinfo {author} {\bibfnamefont {A.}~\bibnamefont
  {Chantasri}}\ and\ \bibinfo {author} {\bibfnamefont {A.~N.}\ \bibnamefont
  {Jordan}},\ }\href@noop {} {\bibfield  {journal} {\bibinfo  {journal} {Phys.
  Rev. A}\ }\textbf {\bibinfo {volume} {92}},\ \bibinfo {pages} {032125}
  (\bibinfo {year} {2015})}\BibitemShut {NoStop}%
\bibitem [{\citenamefont {Chantasri}\ \emph {et~al.}(2013)\citenamefont
  {Chantasri}, \citenamefont {Dressel},\ and\ \citenamefont {Jordan}}]{CWLM4}%
  \BibitemOpen
  \bibfield  {author} {\bibinfo {author} {\bibfnamefont {A.}~\bibnamefont
  {Chantasri}}, \bibinfo {author} {\bibfnamefont {J.}~\bibnamefont {Dressel}},
  \ and\ \bibinfo {author} {\bibfnamefont {A.~N.}\ \bibnamefont {Jordan}},\
  }\href@noop {} {\bibfield  {journal} {\bibinfo  {journal} {Phys. Rev. A}\
  }\textbf {\bibinfo {volume} {88}},\ \bibinfo {pages} {042110} (\bibinfo
  {year} {2013})}\BibitemShut {NoStop}%
\bibitem [{\citenamefont {Hatridge}\ \emph {et~al.}(2013)\citenamefont
  {Hatridge}, \citenamefont {Shankar}, \citenamefont {Mirrahimi}, \citenamefont
  {Schackert}, \citenamefont {Geerlings}, \citenamefont {Brecht}, \citenamefont
  {Sliwa}, \citenamefont {Abdo}, \citenamefont {Frunzio}, \citenamefont
  {Girvin}, \citenamefont {Schoelkopf},\ and\ \citenamefont
  {Devoret}}]{Devoret}%
  \BibitemOpen
  \bibfield  {author} {\bibinfo {author} {\bibfnamefont {M.}~\bibnamefont
  {Hatridge}}, \bibinfo {author} {\bibfnamefont {S.}~\bibnamefont {Shankar}},
  \bibinfo {author} {\bibfnamefont {M.}~\bibnamefont {Mirrahimi}}, \bibinfo
  {author} {\bibfnamefont {F.}~\bibnamefont {Schackert}}, \bibinfo {author}
  {\bibfnamefont {K.}~\bibnamefont {Geerlings}}, \bibinfo {author}
  {\bibfnamefont {T.}~\bibnamefont {Brecht}}, \bibinfo {author} {\bibfnamefont
  {K.~M.}\ \bibnamefont {Sliwa}}, \bibinfo {author} {\bibfnamefont
  {B.}~\bibnamefont {Abdo}}, \bibinfo {author} {\bibfnamefont {L.}~\bibnamefont
  {Frunzio}}, \bibinfo {author} {\bibfnamefont {S.~M.}\ \bibnamefont {Girvin}},
  \bibinfo {author} {\bibfnamefont {R.~J.}\ \bibnamefont {Schoelkopf}}, \ and\
  \bibinfo {author} {\bibfnamefont {M.~H.}\ \bibnamefont {Devoret}},\ }\href
  {\doibase 10.1126/science.1226897} {\bibfield  {journal} {\bibinfo  {journal}
  {Science}\ }\textbf {\bibinfo {volume} {339}},\ \bibinfo {pages} {178}
  (\bibinfo {year} {2013})}\BibitemShut {NoStop}%
\bibitem [{\citenamefont {Murch}\ \emph {et~al.}(2013)\citenamefont {Murch},
  \citenamefont {Weber}, \citenamefont {Macklin},\ and\ \citenamefont
  {Siddiqi}}]{SiddiqiSingle}%
  \BibitemOpen
  \bibfield  {author} {\bibinfo {author} {\bibfnamefont {K.~W.}\ \bibnamefont
  {Murch}}, \bibinfo {author} {\bibfnamefont {S.~J.}\ \bibnamefont {Weber}},
  \bibinfo {author} {\bibfnamefont {C.}~\bibnamefont {Macklin}}, \ and\
  \bibinfo {author} {\bibfnamefont {I.}~\bibnamefont {Siddiqi}},\ }\href@noop
  {} {\bibfield  {journal} {\bibinfo  {journal} {Nature}\ }\textbf {\bibinfo
  {volume} {502}},\ \bibinfo {pages} {211} (\bibinfo {year}
  {2013})}\BibitemShut {NoStop}%
\bibitem [{\citenamefont {Roch}\ \emph {et~al.}(2014)\citenamefont {Roch},
  \citenamefont {Schwartz}, \citenamefont {Motzoi}, \citenamefont {Macklin},
  \citenamefont {Vijay}, \citenamefont {Eddins}, \citenamefont {Korotkov},
  \citenamefont {Whaley}, \citenamefont {Sarovar},\ and\ \citenamefont
  {Siddiqi}}]{SiddiqiEntanglement}%
  \BibitemOpen
  \bibfield  {author} {\bibinfo {author} {\bibfnamefont {N.}~\bibnamefont
  {Roch}}, \bibinfo {author} {\bibfnamefont {M.~E.}\ \bibnamefont {Schwartz}},
  \bibinfo {author} {\bibfnamefont {F.}~\bibnamefont {Motzoi}}, \bibinfo
  {author} {\bibfnamefont {C.}~\bibnamefont {Macklin}}, \bibinfo {author}
  {\bibfnamefont {R.}~\bibnamefont {Vijay}}, \bibinfo {author} {\bibfnamefont
  {A.~W.}\ \bibnamefont {Eddins}}, \bibinfo {author} {\bibfnamefont {A.~N.}\
  \bibnamefont {Korotkov}}, \bibinfo {author} {\bibfnamefont {K.~B.}\
  \bibnamefont {Whaley}}, \bibinfo {author} {\bibfnamefont {M.}~\bibnamefont
  {Sarovar}}, \ and\ \bibinfo {author} {\bibfnamefont {I.}~\bibnamefont
  {Siddiqi}},\ }\href {\doibase 10.1103/PhysRevLett.112.170501} {\bibfield
  {journal} {\bibinfo  {journal} {Phys. Rev. Lett.}\ }\textbf {\bibinfo
  {volume} {112}},\ \bibinfo {pages} {170501} (\bibinfo {year}
  {2014})}\BibitemShut {NoStop}%
\bibitem [{\citenamefont {Campagne-Ibarcq}\ \emph {et~al.}(2014)\citenamefont
  {Campagne-Ibarcq}, \citenamefont {Bretheau}, \citenamefont {Flurin},
  \citenamefont {Auff\`eves}, \citenamefont {Mallet},\ and\ \citenamefont
  {Huard}}]{Huard}%
  \BibitemOpen
  \bibfield  {author} {\bibinfo {author} {\bibfnamefont {P.}~\bibnamefont
  {Campagne-Ibarcq}}, \bibinfo {author} {\bibfnamefont {L.}~\bibnamefont
  {Bretheau}}, \bibinfo {author} {\bibfnamefont {E.}~\bibnamefont {Flurin}},
  \bibinfo {author} {\bibfnamefont {A.}~\bibnamefont {Auff\`eves}}, \bibinfo
  {author} {\bibfnamefont {F.}~\bibnamefont {Mallet}}, \ and\ \bibinfo {author}
  {\bibfnamefont {B.}~\bibnamefont {Huard}},\ }\href {\doibase
  10.1103/PhysRevLett.112.180402} {\bibfield  {journal} {\bibinfo  {journal}
  {Phys. Rev. Lett.}\ }\textbf {\bibinfo {volume} {112}},\ \bibinfo {pages}
  {180402} (\bibinfo {year} {2014})}\BibitemShut {NoStop}%
\bibitem [{\citenamefont {Groen}\ \emph {et~al.}(2013)\citenamefont {Groen},
  \citenamefont {Rist\`e}, \citenamefont {Tornberg}, \citenamefont {Cramer},
  \citenamefont {de~Groot}, \citenamefont {Picot}, \citenamefont {Johansson},\
  and\ \citenamefont {DiCarlo}}]{DiCarlo}%
  \BibitemOpen
  \bibfield  {author} {\bibinfo {author} {\bibfnamefont {J.~P.}\ \bibnamefont
  {Groen}}, \bibinfo {author} {\bibfnamefont {D.}~\bibnamefont {Rist\`e}},
  \bibinfo {author} {\bibfnamefont {L.}~\bibnamefont {Tornberg}}, \bibinfo
  {author} {\bibfnamefont {J.}~\bibnamefont {Cramer}}, \bibinfo {author}
  {\bibfnamefont {P.~C.}\ \bibnamefont {de~Groot}}, \bibinfo {author}
  {\bibfnamefont {T.}~\bibnamefont {Picot}}, \bibinfo {author} {\bibfnamefont
  {G.}~\bibnamefont {Johansson}}, \ and\ \bibinfo {author} {\bibfnamefont
  {L.}~\bibnamefont {DiCarlo}},\ }\href {\doibase
  10.1103/PhysRevLett.111.090506} {\bibfield  {journal} {\bibinfo  {journal}
  {Phys. Rev. Lett.}\ }\textbf {\bibinfo {volume} {111}},\ \bibinfo {pages}
  {090506} (\bibinfo {year} {2013})}\BibitemShut {NoStop}%
\bibitem [{\citenamefont {Weber}\ \emph {et~al.}(2014)\citenamefont {Weber},
  \citenamefont {Chantasri}, \citenamefont {Dressel}, \citenamefont {Jordan},
  \citenamefont {Murch},\ and\ \citenamefont {Siddiqi}}]{SiddiqiMapping}%
  \BibitemOpen
  \bibfield  {author} {\bibinfo {author} {\bibfnamefont {S.~J.}\ \bibnamefont
  {Weber}}, \bibinfo {author} {\bibfnamefont {A.}~\bibnamefont {Chantasri}},
  \bibinfo {author} {\bibfnamefont {J.}~\bibnamefont {Dressel}}, \bibinfo
  {author} {\bibfnamefont {A.~N.}\ \bibnamefont {Jordan}}, \bibinfo {author}
  {\bibfnamefont {K.~W.}\ \bibnamefont {Murch}}, \ and\ \bibinfo {author}
  {\bibfnamefont {I.}~\bibnamefont {Siddiqi}},\ }\href@noop {} {\bibfield
  {journal} {\bibinfo  {journal} {Nature}\ }\textbf {\bibinfo {volume} {511}},\
  \bibinfo {pages} {570} (\bibinfo {year} {2014})}\BibitemShut {NoStop}%
\bibitem [{\citenamefont {Tan}\ \emph {et~al.}(2015)\citenamefont {Tan},
  \citenamefont {Weber}, \citenamefont {Siddiqi}, \citenamefont {M\o{}lmer},\
  and\ \citenamefont {Murch}}]{SiddiqiMolmer}%
  \BibitemOpen
  \bibfield  {author} {\bibinfo {author} {\bibfnamefont {D.}~\bibnamefont
  {Tan}}, \bibinfo {author} {\bibfnamefont {S.~J.}\ \bibnamefont {Weber}},
  \bibinfo {author} {\bibfnamefont {I.}~\bibnamefont {Siddiqi}}, \bibinfo
  {author} {\bibfnamefont {K.}~\bibnamefont {M\o{}lmer}}, \ and\ \bibinfo
  {author} {\bibfnamefont {K.~W.}\ \bibnamefont {Murch}},\ }\href {\doibase
  10.1103/PhysRevLett.114.090403} {\bibfield  {journal} {\bibinfo  {journal}
  {Phys. Rev. Lett.}\ }\textbf {\bibinfo {volume} {114}},\ \bibinfo {pages}
  {090403} (\bibinfo {year} {2015})}\BibitemShut {NoStop}%
\bibitem [{\citenamefont {Ruskov}\ \emph {et~al.}(2010)\citenamefont {Ruskov},
  \citenamefont {Korotkov},\ and\ \citenamefont
  {M\o{}lmer}}]{MolmerMonitoring}%
  \BibitemOpen
  \bibfield  {author} {\bibinfo {author} {\bibfnamefont {R.}~\bibnamefont
  {Ruskov}}, \bibinfo {author} {\bibfnamefont {A.~N.}\ \bibnamefont
  {Korotkov}}, \ and\ \bibinfo {author} {\bibfnamefont {K.}~\bibnamefont
  {M\o{}lmer}},\ }\href {\doibase 10.1103/PhysRevLett.105.100506} {\bibfield
  {journal} {\bibinfo  {journal} {Phys. Rev. Lett.}\ }\textbf {\bibinfo
  {volume} {105}},\ \bibinfo {pages} {100506} (\bibinfo {year}
  {2010})}\BibitemShut {NoStop}%
\bibitem [{\citenamefont {Garc\'{\i}a-Pintos}\ and\ \citenamefont
  {Dressel}(2016)}]{ProbingQ}%
  \BibitemOpen
  \bibfield  {author} {\bibinfo {author} {\bibfnamefont {L.~P.}\ \bibnamefont
  {Garc\'{\i}a-Pintos}}\ and\ \bibinfo {author} {\bibfnamefont
  {J.}~\bibnamefont {Dressel}},\ }\href {\doibase 10.1103/PhysRevA.94.062119}
  {\bibfield  {journal} {\bibinfo  {journal} {Phys. Rev. A}\ }\textbf {\bibinfo
  {volume} {94}},\ \bibinfo {pages} {062119} (\bibinfo {year}
  {2016})}\BibitemShut {NoStop}%
\bibitem [{\citenamefont {Hacohen-Gourgy}\ \emph {et~al.}(2016)\citenamefont
  {Hacohen-Gourgy}, \citenamefont {Martin}, \citenamefont {Flurin},
  \citenamefont {Ramasesh}, \citenamefont {Whaley},\ and\ \citenamefont
  {Siddiqi}}]{SiddiqiDynamics}%
  \BibitemOpen
  \bibfield  {author} {\bibinfo {author} {\bibfnamefont {S.}~\bibnamefont
  {Hacohen-Gourgy}}, \bibinfo {author} {\bibfnamefont {L.~S.}\ \bibnamefont
  {Martin}}, \bibinfo {author} {\bibfnamefont {E.}~\bibnamefont {Flurin}},
  \bibinfo {author} {\bibfnamefont {V.~V.}\ \bibnamefont {Ramasesh}}, \bibinfo
  {author} {\bibfnamefont {K.~B.}\ \bibnamefont {Whaley}}, \ and\ \bibinfo
  {author} {\bibfnamefont {I.}~\bibnamefont {Siddiqi}},\ }\href
  {http://dx.doi.org/10.1038/nature19762} {\bibfield  {journal} {\bibinfo
  {journal} {Nature}\ }\textbf {\bibinfo {volume} {538}},\ \bibinfo {pages}
  {491} (\bibinfo {year} {2016})}\BibitemShut {NoStop}%
\bibitem [{\citenamefont {Juan~Atalaya}\ and\ \citenamefont
  {Korotkov}(2017)}]{SiddiqiKorotkovCorrelators}%
  \BibitemOpen
  \bibfield  {author} {\bibinfo {author} {\bibfnamefont {L.~S. M. I.~S.}\
  \bibnamefont {Juan~Atalaya}, \bibfnamefont {Shay Hacohen-Gourgy}}\ and\
  \bibinfo {author} {\bibfnamefont {A.~N.}\ \bibnamefont {Korotkov}},\
  }\href@noop {} {\bibfield  {journal} {\bibinfo  {journal} {arXiv:1702.08077
  [quant-ph]}\ } (\bibinfo {year} {2017})}\BibitemShut {NoStop}%
\bibitem [{\citenamefont {Yamamoto}\ and\ \citenamefont
  {Haus}(1986)}]{SimOptics}%
  \BibitemOpen
  \bibfield  {author} {\bibinfo {author} {\bibfnamefont {Y.}~\bibnamefont
  {Yamamoto}}\ and\ \bibinfo {author} {\bibfnamefont {H.~A.}\ \bibnamefont
  {Haus}},\ }\href {\doibase 10.1103/RevModPhys.58.1001} {\bibfield  {journal}
  {\bibinfo  {journal} {Rev. Mod. Phys.}\ }\textbf {\bibinfo {volume} {58}},\
  \bibinfo {pages} {1001} (\bibinfo {year} {1986})}\BibitemShut {NoStop}%
\bibitem [{\citenamefont {Aharonov}\ \emph
  {et~al.}(1988{\natexlab{a}})\citenamefont {Aharonov}, \citenamefont
  {Albert},\ and\ \citenamefont {Vaidman}}]{BigWeakValues}%
  \BibitemOpen
  \bibfield  {author} {\bibinfo {author} {\bibfnamefont {Y.}~\bibnamefont
  {Aharonov}}, \bibinfo {author} {\bibfnamefont {D.~Z.}\ \bibnamefont
  {Albert}}, \ and\ \bibinfo {author} {\bibfnamefont {L.}~\bibnamefont
  {Vaidman}},\ }\href {\doibase 10.1103/PhysRevLett.60.1351.} {\bibfield
  {journal} {\bibinfo  {journal} {Physical Review Letters}\ }\textbf {\bibinfo
  {volume} {60}},\ \bibinfo {pages} {1351–1354} (\bibinfo {year}
  {1988}{\natexlab{a}})}\BibitemShut {NoStop}%
\bibitem [{\citenamefont {Wiseman}(2002)}]{WisemanWeakValues}%
  \BibitemOpen
  \bibfield  {author} {\bibinfo {author} {\bibfnamefont {H.~M.}\ \bibnamefont
  {Wiseman}},\ }\href {\doibase 10.1103/PhysRevA.65.032111} {\bibfield
  {journal} {\bibinfo  {journal} {Phys. Rev. A}\ }\textbf {\bibinfo {volume}
  {65}},\ \bibinfo {pages} {032111} (\bibinfo {year} {2002})}\BibitemShut
  {NoStop}%
\bibitem [{\citenamefont {Gammelmark}\ \emph {et~al.}(2013)\citenamefont
  {Gammelmark}, \citenamefont {Julsgaard},\ and\ \citenamefont
  {M\o{}lmer}}]{PastQStateMolmer}%
  \BibitemOpen
  \bibfield  {author} {\bibinfo {author} {\bibfnamefont {S.}~\bibnamefont
  {Gammelmark}}, \bibinfo {author} {\bibfnamefont {B.}~\bibnamefont
  {Julsgaard}}, \ and\ \bibinfo {author} {\bibfnamefont {K.}~\bibnamefont
  {M\o{}lmer}},\ }\href {\doibase 10.1103/PhysRevLett.111.160401} {\bibfield
  {journal} {\bibinfo  {journal} {Phys. Rev. Lett.}\ }\textbf {\bibinfo
  {volume} {111}},\ \bibinfo {pages} {160401} (\bibinfo {year}
  {2013})}\BibitemShut {NoStop}%
\bibitem [{\citenamefont {Nazarov}\ and\ \citenamefont
  {Kindermann}(2003)}]{NazKin}%
  \BibitemOpen
  \bibfield  {author} {\bibinfo {author} {\bibfnamefont {Y.~V.}\ \bibnamefont
  {Nazarov}}\ and\ \bibinfo {author} {\bibfnamefont {M.}~\bibnamefont
  {Kindermann}},\ }\href {\doibase 10.1140/epjb/e2003-00293-1} {\bibfield
  {journal} {\bibinfo  {journal} {The European Physical Journal B - Condensed
  Matter and Complex Systems}\ }\textbf {\bibinfo {volume} {35}},\ \bibinfo
  {pages} {413} (\bibinfo {year} {2003})}\BibitemShut {NoStop}%
\bibitem [{\citenamefont {Franquet}\ and\ \citenamefont
  {Nazarov}(2017)}]{FranquetNaz}%
  \BibitemOpen
  \bibfield  {author} {\bibinfo {author} {\bibfnamefont {A.}~\bibnamefont
  {Franquet}}\ and\ \bibinfo {author} {\bibfnamefont {Y.~V.}\ \bibnamefont
  {Nazarov}},\ }\href {\doibase 10.1103/PhysRevB.95.085427} {\bibfield
  {journal} {\bibinfo  {journal} {Phys. Rev. B}\ }\textbf {\bibinfo {volume}
  {95}},\ \bibinfo {pages} {085427} (\bibinfo {year} {2017})}\BibitemShut
  {NoStop}%
\bibitem [{\citenamefont {Nazarov}\ and\ \citenamefont
  {Danon}(2013)}]{AQMbook}%
  \BibitemOpen
  \bibfield  {author} {\bibinfo {author} {\bibfnamefont {Y.~V.}\ \bibnamefont
  {Nazarov}}\ and\ \bibinfo {author} {\bibfnamefont {J.}~\bibnamefont
  {Danon}},\ }\href@noop {} {\emph {\bibinfo {title} {Advanced Quantum
  Mechanics: A practical guide}}}\ (\bibinfo  {publisher} {Cambridge University
  Press},\ \bibinfo {year} {2013})\BibitemShut {NoStop}%
\bibitem [{\citenamefont {Wiseman}\ and\ \citenamefont
  {Milburn}(1993)}]{trajectory}%
  \BibitemOpen
  \bibfield  {author} {\bibinfo {author} {\bibfnamefont {H.}~\bibnamefont
  {Wiseman}}\ and\ \bibinfo {author} {\bibfnamefont {G.}~\bibnamefont
  {Milburn}},\ }\href {\doibase 10.1103/PhysRevLett.70.548} {\bibfield
  {journal} {\bibinfo  {journal} {Phys.Rev. Lett.}\ }\textbf {\bibinfo {volume}
  {70}},\ \bibinfo {pages} {548} (\bibinfo {year} {1993})}\BibitemShut
  {NoStop}%
\bibitem [{\citenamefont {Clerk}\ \emph {et~al.}(2010)\citenamefont {Clerk},
  \citenamefont {Devoret}, \citenamefont {Girvin}, \citenamefont {Marquardt},\
  and\ \citenamefont {Schoelkopf}}]{QNoise}%
  \BibitemOpen
  \bibfield  {author} {\bibinfo {author} {\bibfnamefont {A.~A.}\ \bibnamefont
  {Clerk}}, \bibinfo {author} {\bibfnamefont {M.~H.}\ \bibnamefont {Devoret}},
  \bibinfo {author} {\bibfnamefont {S.~M.}\ \bibnamefont {Girvin}}, \bibinfo
  {author} {\bibfnamefont {F.}~\bibnamefont {Marquardt}}, \ and\ \bibinfo
  {author} {\bibfnamefont {R.~J.}\ \bibnamefont {Schoelkopf}},\ }\href
  {\doibase 10.1103/RevModPhys.82.1155} {\bibfield  {journal} {\bibinfo
  {journal} {Rev. Mod. Phys.}\ }\textbf {\bibinfo {volume} {82}},\ \bibinfo
  {pages} {1155} (\bibinfo {year} {2010})}\BibitemShut {NoStop}%
\bibitem [{\citenamefont {Aharonov}\ \emph
  {et~al.}(1988{\natexlab{b}})\citenamefont {Aharonov}, \citenamefont
  {Albert},\ and\ \citenamefont {Vaidman}}]{WeakValues}%
  \BibitemOpen
  \bibfield  {author} {\bibinfo {author} {\bibfnamefont {Y.}~\bibnamefont
  {Aharonov}}, \bibinfo {author} {\bibfnamefont {D.~Z.}\ \bibnamefont
  {Albert}}, \ and\ \bibinfo {author} {\bibfnamefont {L.}~\bibnamefont
  {Vaidman}},\ }\href {\doibase 10.1103/PhysRevLett.60.1351} {\bibfield
  {journal} {\bibinfo  {journal} {Phys. Rev. Lett.}\ }\textbf {\bibinfo
  {volume} {60}},\ \bibinfo {pages} {1351} (\bibinfo {year}
  {1988}{\natexlab{b}})}\BibitemShut {NoStop}%
\end{thebibliography}%

\appendix
\section{Positivity of the distribution}
\label{sec:Positivity}
Here we present the derivation of the inequality \eqref{ineq2} from the analytical procedure used in Sec. \ref{sec:smalltimes} to derive the joint distribution of measurement outcomes at short times ${\cal T}$ and given a vanishing overlap between preparation and post-selection states, $\braket{i}{\Psi}=0$.\\
To do so, we focus first on the simple setup considered in the main text. Next, we add different correlations between the two detectors, understand what they add to the picture and derive a more general restriction.\\

To start with, note that for any pair of operators $\hat{Q},\hat{V}$ it is possible to construct a Cauchy-Schwarz inequality of the following form\cite{QNoise}:

\begin{widetext}

\begin{equation}
S_{QQ}	(\omega)S_{VV}(\omega) - \abs{S_{QV}(\omega)}^2\geq \abs{\frac{\hbar}{2}(a_{VQ}(\omega)-a_{QV}(\omega))}^2\left(1+\Delta\left[\frac{S_{QV}(\omega)}{\frac{\hbar}{2}(a_{VQ}(\omega)-a_{QV}(\omega))}\right]\right)
\end{equation}

\end{widetext}

where 
\begin{equation}
\Delta[z]=[\abs{1+z^2}-(1+\abs{z}^2)]/2.
\end{equation}

In the limit of zero frequency, this reproduces the inequality \eqref{ineqideal}.\\

However, in the case where we have more than one detector or measured variable, i.e., more than one pair of input-output variables $\hat{Q},\hat{V}$, there are additional inequalities restricting the correlators between input-output variables pertaining to these different pairs. An easy way to see this is to inspect the distribution we calculated for short time scales, Eq. \eqref{smalltimes}. It describes the case of  independent detectors without cross noises. Let us find the conditions for it to be positive at all values of ${\cal O}_{1,2}$. This condition reads:
\begin{equation}
\gamma \geq \frac{1}{t_{a1}} + \frac{1}{t_{a2}}.
\end{equation}

For us, the inequality can be written as,

\begin{equation}
\label{App:condition1}
S^{(1,1)}_{QQ} + S^{(2,2)}_{QQ} \geq \frac{\hbar^2}{4}\left(\frac{\abs{a^{(1,1)}_{VQ}}^2}{S^{(1,1)}_{VV}} + \frac{\abs{a^{(2,2)}_{VQ}}^2}{S^{(2,2)}_{VV}}\right).
\end{equation}

We write this assuming the condition of a good amplifier, that is, the direct gain exceeds much the reverse one\cite{QNoise}, $a^{(1,1)}_{VQ}\gg a^{(1,1)}_{QV}$. All results presented here can be extended to a more general situation by replacing $a^{(i,i)}_{VQ}$ with the difference $a^{(i,i)}_{VQ}-a^{(i,i)}_{QV}$.
This inequality can be constructed as the sum of two inequalities of the form \eqref{ineq} for the two sets of input-output variables involved. This fact explicitly shows that this inequality does not add any more restrictions to the correlators than the ones that come from the aforementioned Cauchy-Schwarz inequalities.\\

Now, let us derive the distribution at short time for a more general case where the cross noises and correlations are present, and then check the condition for positivity once again.\\

Firstly, for any correlations between output variables, meaning $S^{(1,2)}_{VV}= S^{(2,1)}_{VV} \neq 0$, the distribution \eqref{smalltimes} will change, however, the condition for positivity will not. In particular, adding correlations between output variables modifies it in the following way,

\begin{widetext}

\begin{equation}
\label{App:outputcorr}
P({\cal O}_1, {\cal O}_2) = \frac{1}{4\gamma + {\cal T}\bar{\Omega}^2}\left(4\gamma+{\cal T}\left[\left(\Omega_x-\frac{4{\cal O}_2}{t_{a2}}-\frac{2{\cal O}_1S^{(1,2)}_{VV}}{a^{(2,2)}_{VQ}a^{(1,1)}_{VQ}}\right)^2+\left(\Omega_y+\frac{4{\cal O}_1}{t_{a1}}+\frac{2{\cal O}_2S^{(1,2)}_{VV}}{a^{(2,2)}_{VQ}a^{(1,1)}_{VQ}}\right)^2-\frac{4}{{\cal T}t_{a2}}-\frac{4}{{\cal T}t_{a1}}\right]\right)P_G({\cal O}_1, {\cal O}_2).
\end{equation}

\end{widetext}

The positivity of the distribution is again guaranteed by the same condition \eqref{App:condition1}.\\
Let us now introduce cross noises between input-output, i.e., $S^{(1,1)}_{QV}, S^{(2,2)}_{QV}, S^{(1,2)}_{QV}, S^{(2,1)}_{QV} \neq 0$. The distribution of measurement outcomes can then be approximated as
\begin{widetext}

\begin{align}
\label{App:cross}
\nonumber
P({\cal O}_1,{\cal O}_2) =& \frac{1}{4\gamma + {\cal T}\bar{\Omega}^2}\biggl[4\gamma  + {\cal T}\biggl(\left(\Omega_x + \left( \frac{2S^{(1,2)}_{QV}}{a^{(2,2)}_{V,Q}}-1\right)\frac{{\cal O}_2}{{\cal T}\sigma^2_2} + \frac{2S^{(1,1)}_{QV}}{a^{(1,1)}_{V,Q}}\frac{{\cal O}_1}{{\cal T}\sigma^2_1}\right)^2+\left(\Omega_y +\left(1 +\frac{2S^{(2,1)}_{QV}}{a^{(1,1)}_{V,Q}}\frac{{\cal O}_1}{{\cal T}\sigma^2_1}\right) +\frac{2S^{(2,2)}_{QV}}{a^{(2,2)}_{V,Q}}\frac{{\cal O}_2}{{\cal T}\sigma^2_2}\right)^2\\
&-\left(1-\frac{2S^{(1,2)}_{QV}}{a^{(2,2)}_{V,Q}}\right)^2\frac{1}{{\cal T}^2\sigma^2_2}-\left(\frac{2S^{(1,1)}_{QV}}{a^{(1,1)}_{V,Q}}\right)^2\frac{1}{{\cal T}^2\sigma^2_1}-\left(1+\frac{2S^{(2,1)}_{QV}}{a^{(1,1)}_{V,Q}}\right)^2\frac{1}{{\cal T}^2\sigma^2_1}-\left(\frac{2S^{(2,2)}_{QV}}{a^{(2,2)}_{V,Q}}\right)^2\frac{1}{{\cal T}^2\sigma^2_2}\biggr)\biggr]P_G({\cal O}_1, {\cal O}_2).
\end{align}

\end{widetext}
Here, $\sigma^2_i = t_{ai}/ 4{\cal T}$.\\

For this distribution to be positive we have the following condition,

\begin{widetext}

\begin{equation}
\gamma -\left(1-\frac{2S^{(1,2)}_{QV}}{a^{(2,2)}_{V,Q}}\right)^2\frac{1}{t_{a2}}-\left(\frac{2S^{(1,1)}_{QV}}{a^{(1,1)}_{V,Q}}\right)^2\frac{1}{t_{a1}}-\left(1+\frac{2S^{(2,1)}_{QV}}{a^{(1,1)}_{V,Q}}\right)^2\frac{1}{t_{a1}}-\left(\frac{2S^{(2,2)}_{QV}}{a^{(2,2)}_{V,Q}}\right)^2\frac{1}{t_{a2}}\geq 0;
\end{equation}

\end{widetext}

which one can write as
\begin{widetext}

\begin{equation}
\label{App:condition2.1}
S^{(1,1)}_{QQ} + S^{(2,2)}_{QQ} \geq \frac{\hbar^2}{4}\left[\left(\left(a^{(2,2)}_{V,Q}-2S^{(1,2)}_{QV}\right)^2+\left(2S^{(2,2)}_{QV}\right)^2\right)\frac{1}{S^{(2,2)}_{VV}}+\left(\left(a^{(1,1)}_{V,Q}+2S^{(2,1)}_{QV}\right)^2+\left(2S^{(1,1)}_{QV}\right)^2\right)\frac{1}{S^{(1,1)}_{V,V}}\right].
\end{equation}

\end{widetext}

Conversely, if one takes the initial state to be $\ket{Z^-}$ and the final projection to be $\ket{Z^+}$, then the condition becomes:

\begin{widetext}

\begin{equation}
\label{App:condition2.2}
S^{(1,1)}_{QQ} + S^{(2,2)}_{QQ} \geq \frac{\hbar^2}{4}\left[\left(\left(a^{(2,2)}_{V,Q}+2S^{(1,2)}_{QV}\right)^2+\left(2S^{(2,2)}_{QV}\right)^2\right)\frac{1}{S^{(2,2)}_{VV}}+\left(\left(a^{(1,1)}_{V,Q}-2S^{(2,1)}_{QV}\right)^2+\left(2S^{(1,1)}_{QV}\right)^2\right)\frac{1}{S^{(1,1)}_{V,V}}\right].
\end{equation}

\end{widetext}

The probability distribution of measurement outcomes should remain positive regardless of the initial and final conditions. Thus, both these inequalities \eqref{App:condition2.1} and \eqref{App:condition2.2}, have to be fulfilled. Taking this into account, one can write the inequality \eqref{ineq2}, where inverse susceptibilities are added back, owing to the possibility of bad amplifiers.\\
This shows that the existence of cross noises between input-output of different detectors imposes a stronger restriction on the possible values for the set of noises and response functions defining a measurement scenario than the usual Cauchy-Schwarz inequalities considered. It is worth noting that we did not consider non-vanishing responses between input-output of different detectors. The analysis can be extended to this case with even more complex restrictions on the correlators for the positivity of the distribution of measurement outcomes.

\end{document}